\documentclass[]{rmaa}

			\usepackage{amsmath}
			\usepackage{pdflscape}

			\usepackage{paralist}

			\usepackage{psfrag,color}

			\usepackage[latin1]{inputenc}

			\usepackage{color}
			\usepackage{longtable}

			\title{Testing the Planet-Metallicity Correlation in M-dwarfs with Gemini GNIRS Spectra} 

			\author{
			  M. J. Hobson,\altaffilmark{1} 
			  E. Jofré,\altaffilmark{1,2}
			  L. García,\altaffilmark{1}
			  R. Petrucci,\altaffilmark{1,2}
			  and M. Gómez\altaffilmark{1,2}}

			\altaffiltext{1}{Observatorio Astronómico de Córdoba, Argentina.}

			\altaffiltext{2}{CONICET, Argentina.}

			\shortauthor{Hobson et al.}
			\shorttitle{Planet-Metallicity Correlation in M-dwarfs}

			\fulladdresses{
			\item L. Garc\'ia, M. G\'omez, M. J. Hobson, E. Jofr\'e, R. Petrucci: Observatorio
			Astronómico de Córdoba, Laprida 854, Córdoba, Argentina (melissa.hobson@lam.fr).
			\item M. G\'omez, E. Jofr\'e, R. Petrucci: Consejo Nacional de Investigaciones
			Científicas y Técnicas (CONICET), Argentina.}

			\listofauthors{M. J. Hobson, E. Jofré, L.Garc\'ia, R. Petrucci \& M. G\'omez}
			\indexauthor{Hobson, M. J.}
			\indexauthor{Jofré, E.}
			\indexauthor{Garc\'ia, L.}
			\indexauthor{Petrucci, R.}
			\indexauthor{G\'omez, M.}

			\abstract{While the planet-metallicity correlation for FGK main-sequence stars hosting giant planets
			is well established, the results are not so clear for M-dwarf stars, for which precise metallicity
			measurements are not straightforward. However, new techniques using near infrared spectra show
			promising results. Using these, we determine stellar parameters and metallicities for a sample
			of 16 M-dwarf stars, 11 of which host planets, with near-infrared spectra from the Gemini Near-Infrared
			Spectrograph (GNIRS). 
			We find that M-dwarfs with planets are preferentially metal-rich compared
			to those without planets. 
			This result, based on GNIRS spectra, is supported by the analysis of a relatively larger sample of
			M stars with planets (18 in total) and a control sample of 213 M stars without known planets,
			obtained from the catalogue of \citet{T15}.  This, on the one hand, coincides with the trend (not only
			for M- but also for solar-type stars) reported in
			the literature and, on the other hand, demonstrates the utility of GNIRS spectra to obtain reliable
			stellar parameters of M stars. We also find that M dwarfs that harbor giant planets are preferentially more
			metallic than those associated with low-mass planets (Neptune or super-Earth type). This trend
			also agrees with that previously reported for solar-type stars. These results would
			favor the core accretion model for planetary formation.} 

			\resumen{Mientras que la correlación planeta-metalicidad para las estrellas de secuencia principal
			FGK que albergan planetas gigantes está bien establecida, los resultados no son tan claros para las
			estrellas enanas M, para las cuales las deteminaciones de metalicidad son complejas. Sin embargo,
			técnicas nuevas con espectros infrarrojos muestran resultados prometedores. Aplicando estos m\'etodos 
			se determinan parámetros estelares y metalicidades de 16 estrellas enanas M, incluyendo 11 con planetas
			detectados, a partir de espectros infrarrojos de resolución moderada obtenidos con el instrumento
			GNIRS (Gemini Near-Infrared Spectrograph) del telescopio Gemini Norte.
			Se encuentra que las enanas M con planetas son preferentemente ricas en metales en
			comparación con aquéllas sin planetas. Este resultado, basado en 
			espectros GNIRS, es avalado por el an\'alisis de una muestra relativamente mayor de
			estrellas M con planetas (18 en total) y una muestra de control de 213 estrellas M sin planetas
			detectados, obtenidas del cat\'alogo de \citet{T15}. Esto, por un lado, coincide con la tendencia
			reportada en la literatura (tanto para estrellas M como de masa solar) y, por el otro, demuestra la utilidad 
			de los espectros de GNIRS para la obtenci\'on de par\'ametros estelares confiables de estrellas M.
			Adicionalmente encontramos que las enanas M que albergan planetas gigantes son 
			preferentemente m\'as met\'alicas que aqu\'ellas asociadas con planetas de menor masa
			(de tipo Neptuno o s\'uper-Tierra). Esta tendencia tambi\'en concuerda con la reportada previamente
			para estrellas de tipo solar. Estos resultados favorecer\'\i an el modelo de
			acreci\'on del n\'ucleo para la formaci\'on planetaria.}

			\addkeyword{Methods: observational}
			\addkeyword{Techniques: spectroscopic}
			\addkeyword{Stars: abundances}
			\addkeyword{Planets and satellites: general}

\begin{document}
			\maketitle

			\section{Introduction}
			\label{intro}

			M-dwarf stars are the greatest stellar component of the Galaxy, representing around $70\%$ of stars \citep{IMF}. They are small, cool, faint stars, with masses ranging from $0.075 M_\odot$ to $0.6 M_\odot$, radii from $0.08 R_\odot$ to $0.62 R_\odot$, temperatures from $2100$ K to $3800$ K, and luminosities from $0.001 L_\odot$ to $0.08 L_\odot$ \citep{Kalt}. As the greatest stellar population, they may also represent the greatest population of planetary hosts (e.g. \citealt{Lada}).

			Nowadays it is widely accepted that FGK main-sequence and subgiant stars hosting gas giant planets are, on average, more metal-rich than stars without detected planets (e.g. \citealt{Gonzalez}, \citealt{FischerValenti}, \citealt{GhezziA}). However, the existence of a planet-metallicity correlation is not so clear for stars hosting Neptune-sized
			and smaller planets (e.g. \citealt{Sousa}, \citealt{Mayor}, \citealt{Neves}, \citealt{Buchhave}, \citealt{Wang}).
			Moreover, the results for giant stars with planets have been ambiguous, and the issue is still debated (e.g. \citealt{Maldonado}, \citealt{Jofre}, \citealt{Reffert}).

			There are two generally accepted models for planetary formation: gravitational instability and core accretion. In the gravitational instability model, dust particles settle into a thin disk with local overdense regions, which are unstable and undergo gravitational collapse. Successive collapses and collisions are responsible for the formation of planets (e.g. \citealt{Goldreich}, \citealt{Boss}, \citealt{Youdin}). In the core accretion model, on the other hand, terrestrial planets and the solid cores of gas planets are formed by the accumulation of planetesimals. If a critical mass is reached before gas depletion, gas accretion from the disk begins and forms a giant planet (e.g. \citealt{Pollack}, \citealt{Mordasini}).

			The planet-metallicity correlation found for main-sequence FGK stars would provide support for
		the core accretion planet formation scenario. A more metallic environment allows the rapid formation of planetary
		cores which can start to accrete gas from the surrounding disk to form gas giant planets \citep{Pollack}, whereas
		in low-metallicity the cores form too slowly for accretion to take place before the disk is depleted (\citealt{coreacc},
		\citealt{Mayor}, \citealt{Mordasini12}). The considerably low metallicity ([Fe/H] $\sim$ $-$0.30 dex, e.g. \citealt{metpoor})
		of several main-sequence (MS) stars hosting giant planets, and the average low metallicity found for giant stars with planets
		($\sim$ $-$0.08 dex; \citealt{Jofre}) raises the issue of giant planet formation within the framework of the
		metallicity-dependent core accretion model. \citet{JohnsonLi} derived the minimum metallicity function required
		for planet formation in the core accretion scenario, and found that none of the $\sim$ 320 planet-hosting stars
		reported by 2011 were below the critical metallicity. Similar results are found for evolved stars with planets
		\citep{Jofre}. On the other hand, it has been suggested that the more massive disks around higher-mass stars,
		such as evolved stars with planets\footnote{Giants with planets usually have masses between
		0.9 and 4 $M_{\odot}$ (e.g., \citealt{Sato}, \citealt{Niedz}, \citealt{Jofre}, \citealt{GhezziJohnson}).},
		would compensate their lower metallic abundances and hence enable giant-planet formation
		(\citealt{Ghezzi}, \citealt{Maldonado}).  The situation for the formation of giant planets
		around M-dwarfs may be even less clear,
		since theoretical predictions within the core accretion model foresee that the giant-planet-formation
		may be inhibited at all radial distances \citep{Laughlin}. 

			Given the role that stellar properties, such as metallicity and mass, may play in the process of planet formation, it is also key to confirm (or refute) the planet-metallicity correlation results for stars at the lower end of the mass scale, such as M dwarfs. However, while for FGK stars metallicities can be determined with great accuracy using high-resolution spectra ($R \gtrsim 30000$) in the optical range, the spectra of M-dwarf stars are extremely complex, with many blended lines and strong molecular bands (e.g., TiO, VO) preventing an easy continuum fit and thus complicating a precise metallicity determination. While some studies (e.g., \citealt{Bean}, \citealt{Woolf}) have attempted determinations using high-resolution optical spectra, they have been limited to few and/or metal-poor stars. \citet{Onehag} adopted a slightly different approach, using high-resolution spectra ($R = 50000$) in the infrared J band; this spectral region has few molecular components, enabling precise continuum and line fitting. The need for high resolution, however, limits this technique, as bright stars and/or long integration times are necessary. Additionally, M dwarfs are typically brighter in the H and K bands than in the J band.
				
				Wide-band photometric calibrations using V$-$K and $M_K$ were performed for M-dwarfs
			by \citet{Bonfils}, \citet{JohnsonApps}, and \citet{Schlaufman}, with conflicting
			results: \citet{JohnsonApps} find that only Jupiter hosts are metal-rich,
			while \citet{Schlaufman} report that both Jupiter and Neptune hosts are metal-rich.
			For Jupiter hosts, both results are therefore in agreement with that found for FGK
			stars, whereas the situation for Neptune hosts is not clear for either FGK stars or M-dwarfs.
			\citet{Bonfils} studied only two planetary hosts, Gl 876 and Gl 436, finding near-solar metallicities for both.
			Additionally, \citet{Schlaufman} note that \citet{Bonfils} systematically underestimate metallicities, while
			\citet{JohnsonApps} overestimate them. The main disadvantage of these photometric techniques is the need for
			absolute magnitudes; their determination requires accurate stellar distances, which limits the stars to
			which these techniques can be applied. However, it is expected that the ongoing ESA astrometric mission,
			Gaia \citep{GAIA} provides accurate distances for all objects down to G $\sim$ 20 mag \footnote{G represents
			the broad-band, white-light, Gaia magnitude with wavelength coverage ~330-1050 nm.}, and therefore
			the number of M stars whose metallicities could be determined by the photometric techniques would be
			significantly increased.
				
			Recently, a new technique for determining the metallicities of M dwarfs via near infrared (nIR) spectra was developed by \citet{RA10} and \citet{T12}. This technique requires only moderate-resolution nIR spectra to reliably estimate metallicities, which greatly reduces the observing time required. Additionally, it is not limited to nearby stars with known parallaxes. It is calibrated using wide FGK-M binaries: assuming a common origin, and therefore a common metallicity, for both stars, the metallicity of the FGK component is measured using high-resolution spectroscopy and assigned to the M-dwarf companion. Linear regressions are then performed between these metallicities and the equivalent widths (EWs) of the selected nIR spectral features, resulting in a best-fit relationship between the EWs and the metallicity. Using this technique, \citet{RA12} find an apparent planet-metallicity correlation over 133 M dwarfs including 11 planet hosts, which is strongest for Hot Jupiters, using spectra from the TripleSpec spectrograph on the Palomar 200-inch Hale Telescope with $R \sim 2700$. Likewise, \citet{T12} find that five giant planet hosts are more metal-rich than four M-dwarf planet hosts without known giant planets, using spectra from the NASA-Infrared Telescope Facility SpeX Spectrograph with $R \sim 2000$ \footnote{We note that H- and/or K-band spectra have not only been applied to derive metallicities of M dwarfs with planets, but also of late-type stars in the Kepler field \citep{Muirhead}, of nearby M-dwarfs \citep{Newton14} and of mid- and late-M dwarfs \citep{Mann2015}, in general.}. 
				
			 \citet{GaidosMann} used JHK spectra to derive metallicities of 121 M dwarfs and study the occurrence of giant planets with metallicity for both M-dwarfs and solar-type stars. Their results hint to a deficiency of giant planets in M dwarfs, although this deficiency is not very significant. More recently, \citet{Gaidos2016}, using a photometric calibration in J$-$H, obtained metallicities for M dwarfs and found that the distribution in metallicity of M dwarfs with planets (usually small planets) is indistinguishable from that of M dwarfs without known planets.
		\citet{Souto}, using high-resolution (R $\sim$ 22,500) H-band spectra from the SDSS-IV-APOGEE survey,
		derived chemical abundances for 13 elements for two M-dwarfs with multiplanetary systems, Kepler 138 and Kepler 186,
		and obtained sub-solar metallicities. They found, however, that in both cases previous metallicity
		determinations from lower-resolution spectra were sub-estimated by 0.1-- 0.2 dex.

				Taking into account the lack of consensus about the planet-metallicity correlation in M dwarf stars,
		in this work we applied the techniques developed in the nIR to homogeneously determine  
		stellar parameters and metallicities of 16 M dwarfs (11 of which have planets) from spectra
		obtained with the Gemini North Near-Infrared Spectrograph (GNIRS). In \S~\ref{obs} we present
		the observations and data reduction; in \S~\ref{analysis} the determinations of the stellar parameters,
		including metallicity; searches for correlations between metallicity and planetary parameters are
		presented in \S~\ref{pmc}. To test if trends suggested by GNIRS spectra are also supported by relatively larger
		samples, we used two subsamples of M dwarfs with and without known planets from the catalogue of \citet{T15}.
		Finally in \S~\ref{conc} we summarize the conclusions.

			\section{Observations and Data Reduction} \label{obs}

			NIR spectra were obtained for a sample of 11 M dwarfs with planets and 5 without known planets using the GNIRS spectrograph, mounted on the Gemini North telescope. The observations were carried out during the 2012B and 2013A semesters (programs GN-2012B-Q-23 and GN-2013A-Q-66 respectively). In the 2012B program, five stars were observed: GJ 176, GJ 179, GJ 250 B, GJ 297.2 B, and GJ 317. The remaining 11 stars (GJ 436, GJ 581, GJ 611 B, GJ 649, GJ 777 B, GJ 783.2 B, GJ 849, GJ 876, GJ 1214, HIP 57050, and HIP 79431) were observed in the 2013A program. The spectrograph was used in cross-dispersed mode, covering a range of $1.2 - 2.5 \ \mu m$, with $R \sim 1700$; to achieve this, the 10.5 l/mm grating, long blue camera (0.05 arcsec/pix) plus the SXD prism and a 0.45 arcsec slit were employed. Table \ref{tab:Mstars} lists the M-stars observed, the V and K magnitudes, the corresponding telluric standard stars, and the number of planets and discovery papers where applicable.

\begin{figure}
			\includegraphics[width=1.15\hsize]{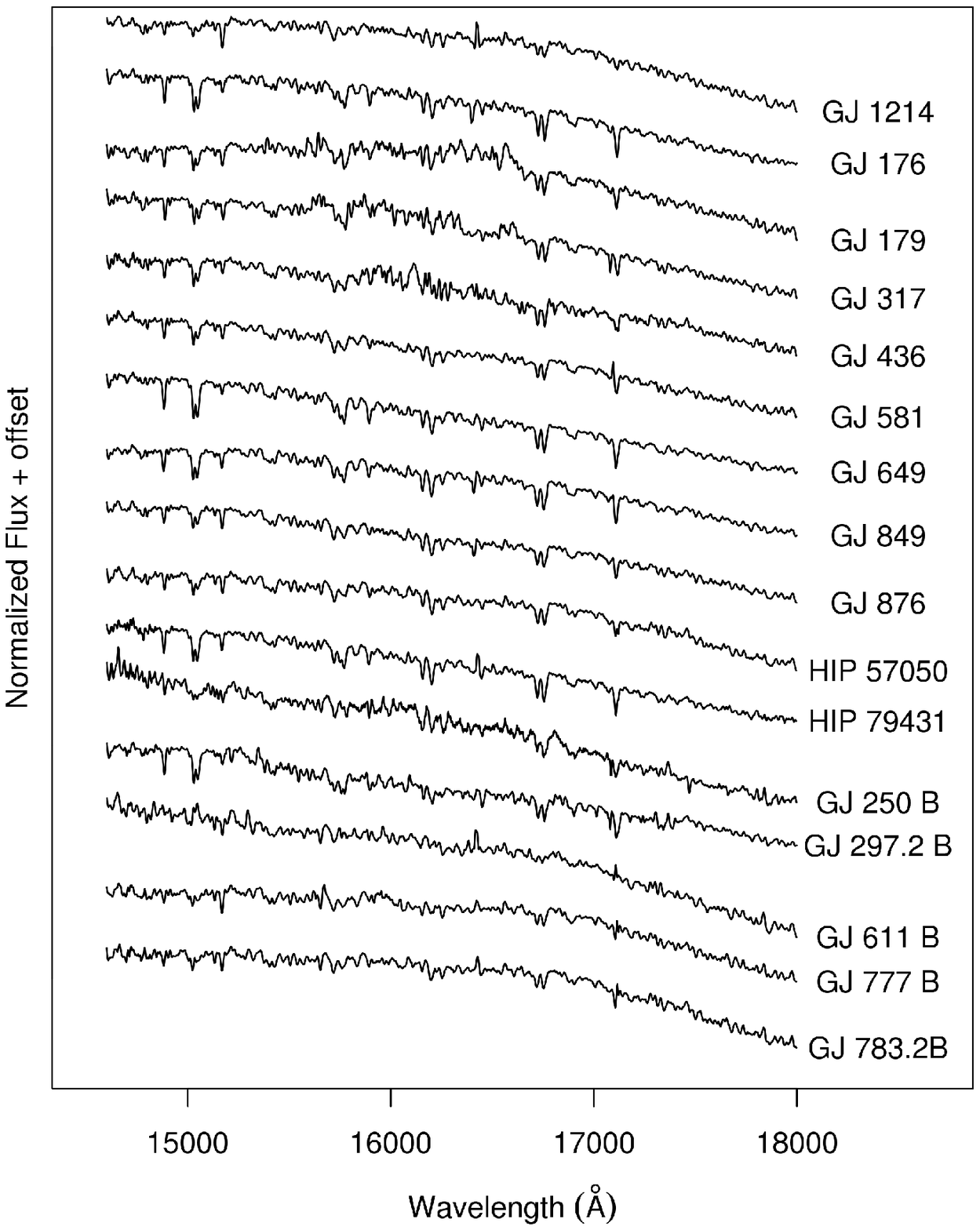}
			\caption{Normalised spectra in the H band. Each spectrum has been normalised by its mean
			flux between 1.46 and 1.80 $\mu$m and arbitrarily shifted.}
			\label{sploth}
			\end{figure}

\begin{figure}
			\includegraphics[width=1.15\hsize]{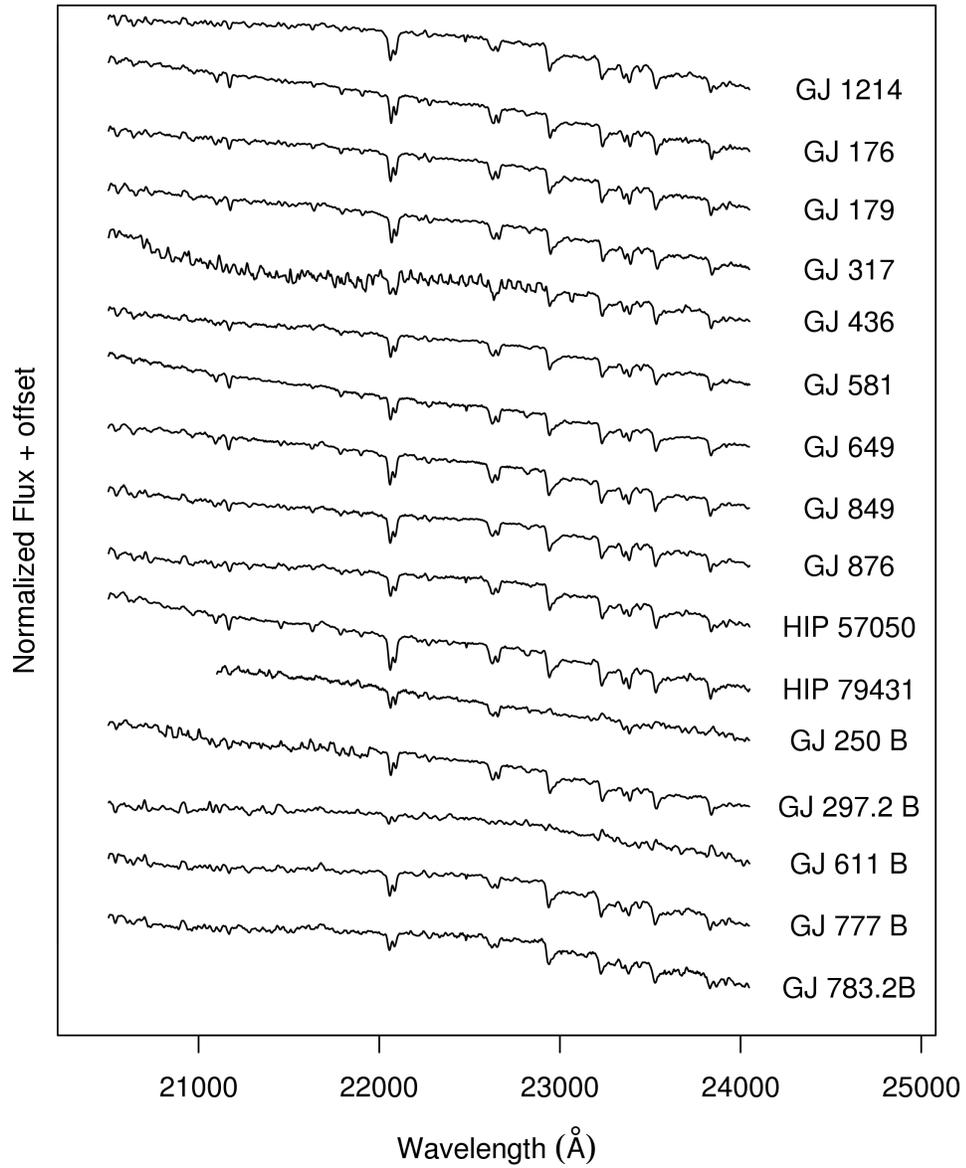}
			\caption{Normalised spectra in the K band. Each spectrum has been normalised by its mean 
			flux between 2.05 and 2.405 $\mu$m and arbitrarily shifted.}
			\label{splotk}
			\end{figure}

			 The spectra were reduced using the XDGNIRS pipeline developed by Rachel Mason and Omaira González-Martín\footnote{Obtained from \url{http://drforum.gemini.edu/wp-content/uploads/2013/11/XDGNIRS_v20.pdf}}. This pipeline creates and subtracts the flatfield, removes electronic pattern noise, cuts and straightens the orders, wavelength calibrates, extracts the spectrum, removes telluric lines (based on a telluric standard, listed in Table \ref{tab:Mstars} for each star), flux calibrates, and calculates the SNR of the spectrum. Finally, the reduced spectra were normalised using standard IRAF\footnote{IRAF is distributed by the National Optical Astronomy Observatories, which are operated by the Association of Universities for Research in Astronomy, Inc., under cooperative agreement with the National Science Foundation.} tasks. The Signal-to-Noise Ratio (SNR) in the observed spectra for the $2.1-2.2 \ \mu m$ region (calculated as mean/rms of the normalised spectra in the region) are listed in Table \ref{tab:Mstars}. Figures \ref{sploth} and \ref{splotk} show the normalised spectra in the H and K bands respectively. The targets in our sample have radial velocities $<$ 100 kms$^{-1}$, so any spectral-feature-wavelength displacement would be $\lesssim$ 10$^{-3}$ $\mu$m, and thus indistinguishable within our resolution.

			\begin{table}
			\centering
			\setlength{\tabnotewidth}{1\columnwidth}
			  \tablecols{6}
			\caption{Observed M-stars with and without planets}
			\label{tab:Mstars}
			\begin{tabular}{p{0.18\linewidth} p{0.08\linewidth} p{0.08\linewidth} p{0.08\linewidth} p{0.18\linewidth} p{0.1\linewidth} p{0.21\linewidth}}
			\toprule
			Star       & V\tabnotemark{$\dagger$} & K\tabnotemark{$\dagger$} & SNR\tabnotemark{$\ddagger$} & Telluric Standard & Planets & Reference     \\ 
			\midrule
			GJ 1214    & 14.67                         & 8.782                         & 101                             & HIP 87108 & 1       & 1             \\
			GJ 176     & 9.951                         & 5.607                         & 85                             & HIP 22913 & 1       & 2             \\
			GJ 179     & 12.018                        & 6.942                         & 95                             & HIP 22913 & 1       & 3             \\
			GJ 317     & 11.975                        & 7.028                         & 83                             & HIP 43269 & 2       & 4             \\
			GJ 436     & 10.613                        & 6.073                         & 40                             & HIP 57014 & 1       & 5             \\
			GJ 581     & 10.560                        & 5.837                         & 91                             & HIP 73249 & 3       & 6, 7, 8       \\
			GJ 649     & 9.655                         & 5.624                         & 86                             & HIP 85790 & 2       & 9, 10         \\
			GJ 849     & 10.366                        & 5.594                         & 86                             & HIP 109442 & 2       & 11, 12        \\
			GJ 876     & 10.192                        & 5.010                         & 96                             & HIP 115119 & 4       & 13, 14, 15, 16 \\
			HIP 57050  & 11.959                        & 6.822                         & 82                             & HIP 57014 & 1       & 17            \\
			HIP 79431  & 11.372                        & 6.589                         & 64                             & HIP 79124 & 1       & 18            \\ \midrule
			Star       & V\tabnotemark{$\dagger$} & K\tabnotemark{$\dagger$} & SNR\tabnotemark{$\ddagger$} & Telluric Standard &         &               \\ \midrule
			GJ 250 B   & 10.05                         & 5.72                          & 42                             & HIP 33420 &         &               \\
			GJ 297.2 B & 11.815                        & 7.418                         & 52                             & HIP 42444 &         &               \\
			GJ 611 B   & 14.206                        & 9.159                         & 82                             & HIP 78649 &         &               \\
			GJ 777 B   & 14.40                         & 8.712                         & 76                             & HIP 98699 &         &               \\
			GJ 783.2 B & 13.932                        & 8.883                         & 67                             & HIP 99742 &         &               \\
			\bottomrule
			\tabnotetext{$\dagger$}{Obtained from SIMBAD, \url{http://simbad.u-strasbg.fr/simbad/} \citep{SIMBAD}.}
			\tabnotetext{$\ddagger$}{SNR for the $2.1-2.2 \ \mu m$ region.}
			\tabnotetext{}{References: (1) \citet{GJ1214};
			(2) \citet{GJ176}; (3) \citet{GJ179}; (4) \citet{GJ317};
			(5) \citet{GJ436}; (6) \citet{GJ581b}; (7) \citet{GJ581c};
			(8) \citet{GJ581e}; (9) \citet{GJ649}; (10) \citet{GJ649c};
			(11) \citet{GJ849}; (12) \citet{GJ849c}; (13) \citet{GJ876b};
			(14) \citet{GJ876c}; (15) \citet{GJ876d}; (16) \citet{GJ876e};
			(17) \citet{HIP57050}; (18) \citet{HIP79431}.}
			\end{tabular}
			\end{table}

			\section{Analysis} \label{analysis}

			\subsection{Fundamental Stellar Parameters} \label{starpar}

			In this section we use our GNIRS/Gemini spectra and the calibrations from
			\citet{Newton} to calculate temperatures, radii, and Log L of the stars in our
			sample and compare them with previous estimations from the literature. The aim is to
			verify whether GNIRS/Gemini spectra are suitable to derive fundamental stellar parameters.

			Calibrations from \citet{Newton} make use of the equivalent widths (EWs) of the
			Al doublet at $1.67 \ \mu m$ and three Mg lines at $1.50$, $1.57$, and $1.71 \ \mu m$ in
			the H band. For details about the precise feature windows and regions used to estimate
			the continuum for each line, we refer the reader to Table 1 of \citet{Newton}. The
			EWs{\footnote{For all
                        EWs measured in this work, 
                        we visually verified that spectral features were included in the defined windows in spite of any
                        (modest) shifts due to stellar radial velocities.}} were calculated from the standard definition:

			\begin{equation} \label{EWeq}
			EW_\lambda = \sum_i \left[ 1-\frac{I(\lambda_i)}{I_c(\lambda_i)}\right]  \Delta \lambda_i ,
			\end{equation}

			\noindent with $i$ the pixels spanned by the line, $\lambda_i$ the wavelength at pixel $i$, $I(\lambda_i)$ the observed line intensity at pixel $i$, $I_c(\lambda_i)$ the calculated continuum intensity at pixel $i$, and $\Delta \lambda_i$ the pixel width.
			The continuum for each line was defined, in accordance with \citet{Newton}, as a linear fit to the corresponding blue and red continuum regions. Once the EWs were measured, we calculated the stellar parameters using the expresions:

			\begin{multline}
			T_{eff}/K = 271.4 \times EW_{Al\text{-}a(1.67\mu m)} \\ + 392.7 \times EW_{Mg(1.50 \mu m)}/EW_{Al\text{-}b(1.67 \mu m)} + 2427
			\end{multline}

			\begin{multline}
			R/R_\odot = -0.0489 \times EW_{Mg(1.57 \mu m)} + 0.275 \times EW_{Al\text{-}a(1.67\mu m)} \\ + 0.201 \times EW_{Mg(1.57 \mu m)}/EW_{Al\text{-}a(1.67 \mu m)} - 0.216
			\end{multline}

			\begin{multline}
			log L/L_\odot = 0.832 \times EW_{Mg(1.71 \mu m)} - 0.176 \times [EW_{Mg(1.71 \mu m)}^2] \\+ 0.266 \times EW_{Mg(1.50 \mu m)} - 3.491,
			\end{multline}
			with $EW_X$ the equivalent width measured for element X.

			The results are reported in Table \ref{tab:starparam}. The errors were estimated using Monte Carlo methods: random Gaussian noise based on the SNR was added to each spectrum, and the EWs were recalculated. We performed this 100 times for each target, and estimated the EW errors as the standard deviation of the 100 runs. The errors of the parameters were then obtained by propagation in quadrature.   All the stellar parameters of GJ 611 B and GJ 1214 were outside the ranges in which \citet{Newton}'s calibrations are valid, so those objects were omitted from the table. Additionally, for other stars only some of the parameters were outside these ranges; for these stars, we report only the parameters with the ranges of validity.

			For the temperatures we found average differences of 13, 34, and 20 K with the estimations from \citet{Mann13},
			\citet{T15}, and \citet{Newton}, respectively. These differences are within the uncertainties of our temperatures.
			With \citet{RA12} we found an average difference of $\sim$ 100 K.
			We point out that this comparison is only possible for 8 and 5 stars in the case of \citet{Mann13}
			and \citet{Newton}, respectively. With regard to \citet{RA12} and \citet{T15} we have 11 stars in common for which temperatures are provided. 

			The comparison of radii and luminosities is rather limited due to the small number of objects.
			However, for the 5 stars (GJ 176, GJ 436, GJ 581,
			GJ 649, and GJ 876) we have in common with \citet{Newton}, we found a reasonably good agreement 
		with an average difference of 0.05 R$_{\odot}$, which is of the same order as the errors in our
		determinations. A similar result was found for Log L with an average difference of 0.1 L$_\odot$.
                \citet{T15} provide radii for 13 stars in common with our sample. Our determinations agree  with thier published 
                values, with an average difference of 0.02 R$_{\odot}$.  Unfortunaltely, no individual values of radius
                and luminosity are reported in the other works cited above. 

	To explore the existence of any systematic difference that might be masked in the determination of
        the final values of 
        temperatures, radii and luminosities, we directly compared our measured EWs for Al-a, Al-b,
        Mg($1.5~\mu\rm m$), Mg($1.57~\mu\rm m$), and Mg($1.71~\mu\rm m$) with those from the literature.
        As mentioned above we have only 5 stars in common with \citet{Newton} 
        which prevents a detailed analysis, and 16 stars with \citet{T15}.
        However, we found a good agreement in both cases, suggesting an absence
        of systematic differences in the EWs measured from GNIRS spectra with respect to previous determinations.
        A more thorough comparison to definitely rule out any difference would require a larger common sample.

	\begin{table}
	\centering
	\setlength{\tabnotewidth}{1\columnwidth}
	  \tablecols{5}
	\caption{Stellar parameters derived in this work for the observed M-stars}
	\label{tab:starparam}
	\begin{tabular}{lrrrr}
	\toprule
	Star       & Temperature {[}K{]} & Radius {[}$R_\odot${]} & Log L {[}$L_\odot${]} \\ \midrule
	GJ 176     & $3531 \pm 50$             & $0.44 \pm 0.04$       & $-1.48 \pm 0.11$        \\
	GJ 179     & $3362 \pm 49$             & $0.40 \pm 0.03$       & $-1.94 \pm 0.11$    \\
	GJ 250 B   & \tabnotemark{$\dagger$}                   & \tabnotemark{$\dagger$}               & $-2.26 \pm 0.22$   \\
	GJ 297.2 B & $3553 \pm 114$            & $0.37 \pm 0.07$       & $-1.54 \pm 0.21$    \\
	GJ 317     & $3234 \pm 45$             & $0.41 \pm 0.04$       & $-2.02 \pm 0.11$        \\
	GJ 436     & $3603 \pm 182$            & $0.31 \pm 0.08$       & $-1.89 \pm 0.20$    \\
	GJ 581     & $3357 \pm 57$             & $0.30 \pm 0.04$       & $-2.06 \pm 0.10$ \\
	GJ 649     & $3695 \pm 51$             & $0.46 \pm 0.03$       & $-1.32 \pm 0.13$  \\
	GJ 777 B   & \tabnotemark{$\dagger$}                   & $0.23 \pm 0.05$       & \tabnotemark{$\dagger$}                           \\
	GJ 783.2 B & \tabnotemark{$\dagger$}                   & $0.23 \pm 0.06$       & \tabnotemark{$\dagger$}                 \\
	GJ 849     & $3408 \pm 45$             & $0.41 \pm 0.04$       & $-1.60 \pm 0.13$       \\
	GJ 876     & $3285 \pm 48$             & $0.29 \pm 0.04$       & $-2.10 \pm 0.10$   \\
	HIP 57050  & $3295 \pm 50$             & $0.30 \pm 0.03$       & $-2.15 \pm 0.10$   \\
	HIP 79431  & $3391 \pm 52$             & $0.42 \pm 0.04$       & $-1.64 \pm 0.15$  \\ \bottomrule
	\tabnotetext{$\dagger$}{Parameter outside the ranges in which \citet{Newton}'s calibrations are valid.}
	\end{tabular}
	\end{table}

	\subsection{Stellar Metallicities} \label{starmet}

\begin{landscape}
        \begin{figure}
        \includegraphics[width=\columnwidth]{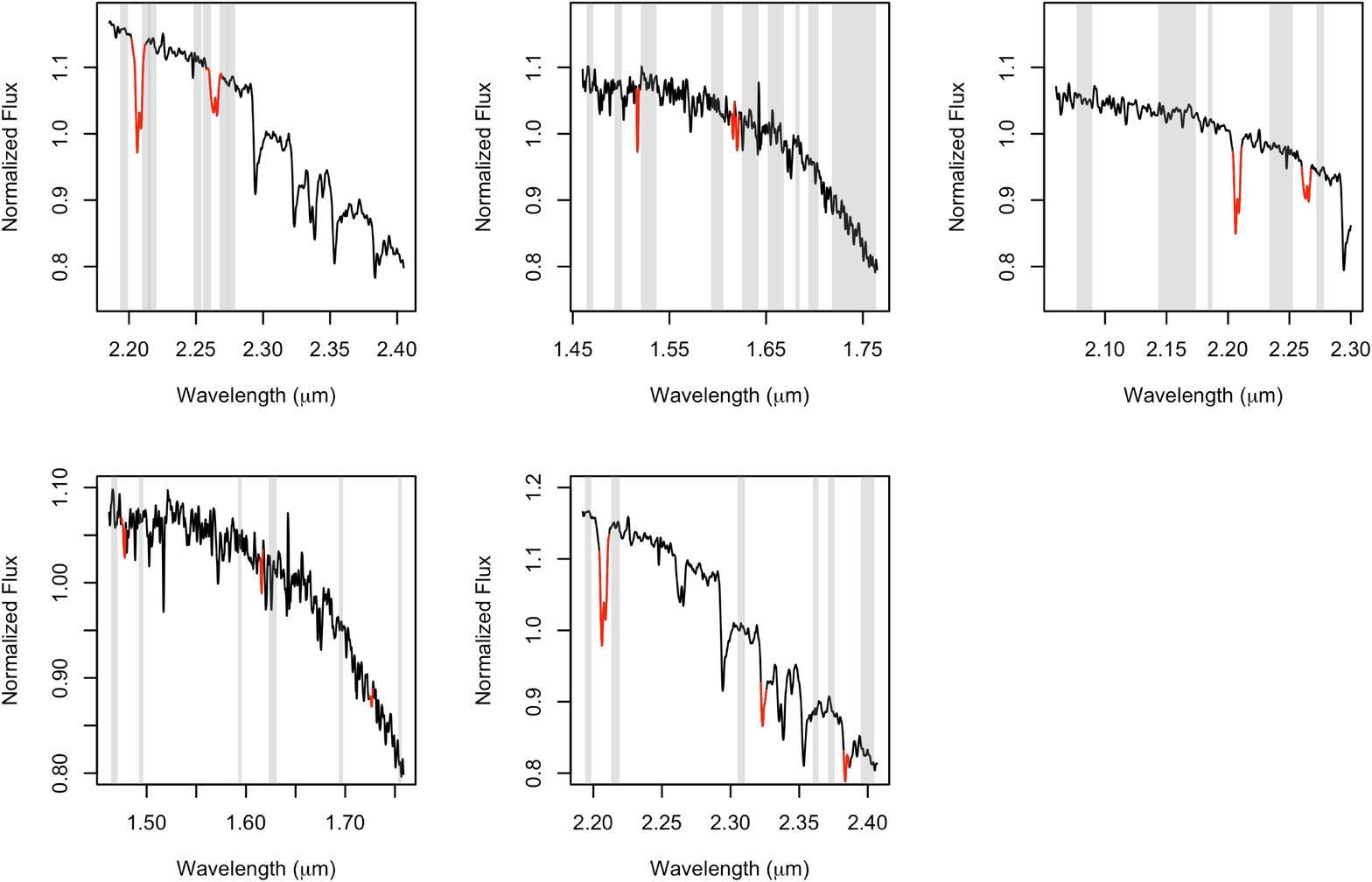}
        \caption{Continuum regions (grey rectangles) and spectral lines (red continuous lines) used to determine metallicity
        for GJ 1214 following: top panels from left to right, \citet{RA12}'s method, \citet{T12}'s method in the H band,
        and \citet{T12}'s method in the K band; bottom panels from left to right, \citet{Mann13}'s method in the H
        and in the K bands, respectively. Each spectrum has been normalised by the mean flux in the shown
        spectral region.}
        \label{contslines}
        \end{figure}
        \end{landscape}

	To estimate the stellar metallicities, we used the calibrations developed by \citet{RA10}, which were updated in \citet{RA12},
	those determined by \citet{T12}, and those developed by \citet{Mann13}. While all these calibrations are based on moderate-resolution nIR spectra, there are differences between them: \citet{RA12} only use lines from the K-band - Na I doublet ($\lambda=2.206$ and $\lambda=2.2069 \ \mu m$) and Ca I triplet ($\lambda=2.261$, $\lambda=2.263$, and $\lambda=2.265 \ \mu m$) - and define the continuum via linear fits to regions close to each line. \citet{T12} give calibrations for both the K-band - Na ($\lambda=2.2074 \ \mu m$) and Ca ($\lambda=2.2638 \ \mu m$) lines - and the H band - Ca ($\lambda=1.6159$ and $\lambda=1.6203 \ \mu m$) and K ($\lambda = 1.5171 \ \mu m$) lines, and define a continuum for the whole of each band using
fourth-order Legendre polynomials. Finally, \citet{Mann13} select regions empirically determined to be sensitive to metallicity changes in the H, J, and K bands, and define the continuum by linear fits over regions close to each line. The precise feature windows for each line and the regions used to define the continuum can be consulted in the corresponding works.  As an example, Figure \ref{contslines} shows the spectral lines employed for each calibration and the regions used to define the continuums for GJ 1214.

	For all calibrations, the EWs were calculated by definition using Eq. \ref{EWeq}. The fourth-order Legendre polynomial continuum fits for \citet{T12}'s calibration were performed using the IRAF \textit{splot} task, whereas the linear continuum fits for \citet{RA12}'s and \citet{Mann13}'s calibrations were carried out with least-square regression methods.

	In addition to the EWs of the spectral lines, all calibrations also use $\mathrm{H_2 O}$ indices in their metallicity determinations to account for the effects of stellar temperature. These indices are given by:

	\begin{equation}
	H_2 O\text{-}K_{RA} = \frac{\langle F(2.070-2.090) \rangle / \langle F(2.235-2.255) \rangle}{\langle F(2.235-2.255) \rangle / \langle F(2.360-2.380) \rangle}
	\end{equation} \citep{RA12} and

	\begin{multline}
	\begin{aligned}
	H_2 O\text{-}K_{T} = \frac{\langle F(2.180-2.200) \rangle / \langle F(2.270-2.290) \rangle}{\langle F(2.270-2.290) \rangle / \langle F(2.360-2.380) \rangle}\\
	H_2 O\text{-}H_{T} = \frac{\langle F(1.595-1.615) \rangle / \langle F(1.680-1.700) \rangle}{\langle F(1.680-1.700) \rangle / \langle F(1.760-1.780) \rangle}
	\end{aligned}
	\end{multline} \citep{T12}, with $\langle F(a-b) \rangle$ the mean flux in the range defined by $a$ and $b$ (in $\mu m$). \citet{Mann13} employed \citet{T12}'s H-band index and \citet{RA12}'s K-band index. The final calibration equations are:

	\begin{multline}
	[Fe/H]_{RA} = -1.039 + 0.092 \times EW_{Na}/H_2O\text{-}K \\ + 0.119 \times EW_{Ca}/H_2 O\text{-}K
	\end{multline} \citep{RA12}

	\begin{multline}
	\begin{aligned}
	\left[Fe/H\right]_{T, K band} = 0.132 \times EW_{Na} + 0.083 \times EW_{Ca} \\
	- 0.403 \times H_2 O\text{-}K - 0.616\\
	[Fe/H]_{T, H band} = 0.340 \times EW_{Ca} + 0.407 \times EW_{K} \\+ 0.436 \times H_2O\text{-}H - 1.485
	\end{aligned}
	\end{multline} \citep{T12}, and

	\begin{multline}
	\begin{aligned}
	\left[Fe/H\right]_{M, K band} = 0.19 \times EW_{F19} + 0.069 \times EW_{F22} + 0.083 \times EW_{F20} \\
	+ 0.218 \times H_2 O\text{-}K - 1.55\\
	[Fe/H]_{M, H band} = 0.40 \times EW_{F17} + 0.51 \times EW_{F14} - 0.28 \times EW_{F18} \\- 1.460 \times H_2O\text{-}H +0.71
	\end{aligned}
	\end{multline} \citep{Mann13}.

	Table \ref{tab:met} presents the metallicities obtained by each method. Both the errors of the EWs, and the quoted uncertainties of the calibrations employed, are sources of uncertainty in the metallicity determinations. Taking a conservative approach, we chose to adopt for the metallicity error the largest value between that obtained by propagating the EWs and $\mathrm{H_2 O}$ indices errors, and the quoted uncertainty of the calibration employed (0.14 for \citealt{RA12}, 0.12 for \citealt{T12}, 0.07 for \citealt{Mann13} in the H band, and 0.06 for \citealt{Mann13} in the K band), in each case.

	The errors for the EWs and $\mathrm{H_2 O}$ indices required by \citet{RA12}'s and \citet{Mann13}'s
	calibrations were estimated using Monte Carlo methods: random Gaussian noise based on the SNR was added to
	each spectrum, and the EWs and $\mathrm{H_2 O}$ indices were recalculated. We performed this 1000 times
	for each target, and estimated the EWs and $\mathrm{H_2 O}$ indices errors as the standard deviation of
	the 1000 runs. For \citet{T12}'s calibration, the more elaborate continuum fit made Monte Carlo error
	estimations too complex. Therefore, we used an analytic method developed by \citet{Sembach}, which was
	also employed by \citet{T12}.  Two stars in Table \ref{tab:met}, GJ 250 B and GJ 611 B, show different
        metallicity values depending on calibrations or spectral bands used. 
        Our spectrum for GJ 250 B has relatively lower SNR in comparison with the rest of the
        stars in Table \ref{tab:met} (see Table \ref{tab:Mstars}). In the case of GJ 611 B, it was not
        possible to achive completely satisfactory telluric correction, resulting in a relatively noisy spectrum.  
        These facts may explain the differences shown in Table \ref{tab:met}.

        The calibrations of \citet{RA12} and \citet{T12} are based on $\sim$ 20 wide binary stars and are valid
        for M0--M4 dwarfs and near solar metallicities ($-$0.4 $\lesssim$ [Fe/H] $\lesssim$ $+$0.3). Using a larger
        number of calibrators (110 wide binaries), \citet{Mann13} refined these calibrations and expanded the limits of
        validity in metallicity ($-$1.04 $<$ [Fe/H] $<$ $+$0.56) and ranges of spectral types (K5--M6). For all bands,
        including J-band, \citet{Mann13} found it possible to obtain reliable metallicities ($<$0.10 dex), although
        features in the K-band provide the best results.

        Recently, \citet{T15} applied the calibrations of \citet{Mann13} to a large sample of M
        dwarfs and compared the results with the ones obtained from other similar calibrations
        such as the ones of \citet{T12}, \citet{Mann2014}, and \citet{Newton14}, all of which
        were developed with data from the same instrument (IRTF-SpeX, R$\sim$2000).
        Finally, they choose the K-band-calibration derived metallicities of \citet{Mann13} as the preferred
        measure of [Fe/H] for M1--M5 dwarfs. This calibration is not only based on a larger number of
        calibrators and has a wider spectral range of validity, but also provides a better agreement with
        other literature measurements and stability against small radial velocity shifts.

        \begin{landscape}
        \begin{table*}
        \centering
        \setlength{\tabnotewidth}{1\columnwidth}
          \tablecols{6}
        \caption{Stellar metallicities derived in this work}
        \label{tab:met}
        \begin{tabular}{lrrrrr}
        \toprule
        Star          & $[Fe/H]_{RA12}\tabnotemark{1}$  & $[Fe/H]_{T12,(K-band)}\tabnotemark{2}$ & $[Fe/H]_{T12,(H-band)}\tabnotemark{3}$ & $[Fe/H]_{M13,(K-band)}\tabnotemark{4}$ & $[Fe/H]_{M13,(H-band)}\tabnotemark{5}$ \\ \midrule
        GJ 176        & $0.008 \pm 0.14$ & $0.08 \pm 0.12$  & $0.17 \pm 0.13$ & $0.17 \pm 0.06$ & $-0.07 \pm 0.09$   \\
        GJ 179        & $0.06 \pm 0.14$  & $0.17 \pm 0.12$   & $0.36 \pm 0.19$ & $0.26 \pm 0.06$ & $0.04 \pm 0.08$   \\
        GJ 250 B      & $-0.34 \pm 0.14$ & $-0.29 \pm 0.14$  & $-0.004 \pm 0.19$ & $-0.79 \pm 0.08$ & $-0.29 \pm 0.19$   \\
        GJ 297.2 B    & $-0.07 \pm 0.14$  & $-0.10 \pm 0.12$  & $-0.17 \pm 0.16$  & $0.00 \pm 0.06$ & $-0.40 \pm 0.15$  \\
        GJ 317        & $0.08 \pm 0.14$   & $0.24 \pm 0.12$   & $0.21 \pm 0.21$ & $0.25 \pm 0.06$ & $-0.08 \pm 0.09$    \\
        GJ 436        & $-0.23 \pm 0.14$  & $-0.20 \pm 0.12$   & $-0.78 \pm 0.19$ & $-0.21 \pm 0.08$ & $-0.38 \pm 0.19$   \\
        GJ 581        & $-0.15 \pm 0.14$  & $-0.01 \pm 0.12$  & $-0.057 \pm 0.12$ & $-0.07 \pm 0.06$ & $-0.22 \pm 0.09$  \\
        GJ 611 B      & $-0.78 \pm 0.14$  &   \tabnotemark{$\dagger$}           &       \tabnotemark{$\dagger$}   & $-1.20 \pm 0.06$ & $-0.93 \pm 0.10$      \\
        GJ 649        & $-0.05 \pm 0.14$  & $0.01 \pm 0.12$   & $0.11 \pm 0.12$  & $0.00 \pm 0.06$ & $-0.15 \pm 0.09$    \\
        GJ 777 B      & $-0.17 \pm 0.14$  & $-0.04 \pm 0.12$  & $-0.006 \pm 0.16$  & $0.03 \pm 0.06$ & $0.07 \pm 0.11$  \\
        GJ 783.2 B    & $-0.36 \pm 0.14$  & $-0.15 \pm 0.12$  & $-0.21 \pm 0.16$  & $-0.25 \pm 0.07$ & $-0.32 \pm 0.13$     \\
        GJ 849        & $0.19 \pm 0.14$  & $0.31 \pm 0.12$   & $0.43 \pm 0.13$  & $0.44 \pm 0.06$ & $0.12 \pm 0.09$    \\
        GJ 876        & $0.09 \pm 0.14$   & $0.22 \pm 0.12$   & $0.10 \pm 0.13$  & $0.32 \pm 0.06$ & $0.01 \pm 0.08$    \\
        GJ 1214       & $-0.008 \pm 0.14$ & $0.19 \pm 0.12$   & $0.11 \pm 0.12$  & $0.41 \pm 0.06$ & $-0.22 \pm 0.08$    \\
        HIP 57050     & $-0.09 \pm 0.14$  & $0.02 \pm 0.12$    & $-0.04 \pm 0.14$ & $0.05 \pm 0.06$ & $-0.15 \pm 0.12$   \\
        HIP 79431     & $0.37 \pm 0.14$   & $0.58 \pm 0.12$    & $0.20 \pm 0.12$ & $0.66 \pm 0.05$    & $0.20 \pm 0.12$    \\ \bottomrule
        \tabnotetext{1}{Metallicities obtained following \citet{RA12}.}
        \tabnotetext{2}{Metallicities obtained following \citet{T12} for the K band.}
        \tabnotetext{3}{Metallicities obtained following \citet{T12} for the H band.}
        \tabnotetext{4}{Metallicities obtained following \citet{Mann13} for the K band.}
        \tabnotetext{5}{Metallicities obtained following \citet{Mann13} for the H band.}
        \tabnotetext{$\dagger$}{\citet{T12}'s methods could not be applied to this star because the spectra was too noisy to allow for the fitting of the Legendre polynomial continuums.}
        \end{tabular}
        \end{table*}
        \end{landscape}

	\subsection{Comparison with the literature and selected calibration}\label{comparison}

\begin{landscape}
	\begin{figure}
	\includegraphics[width=0.3\columnwidth]{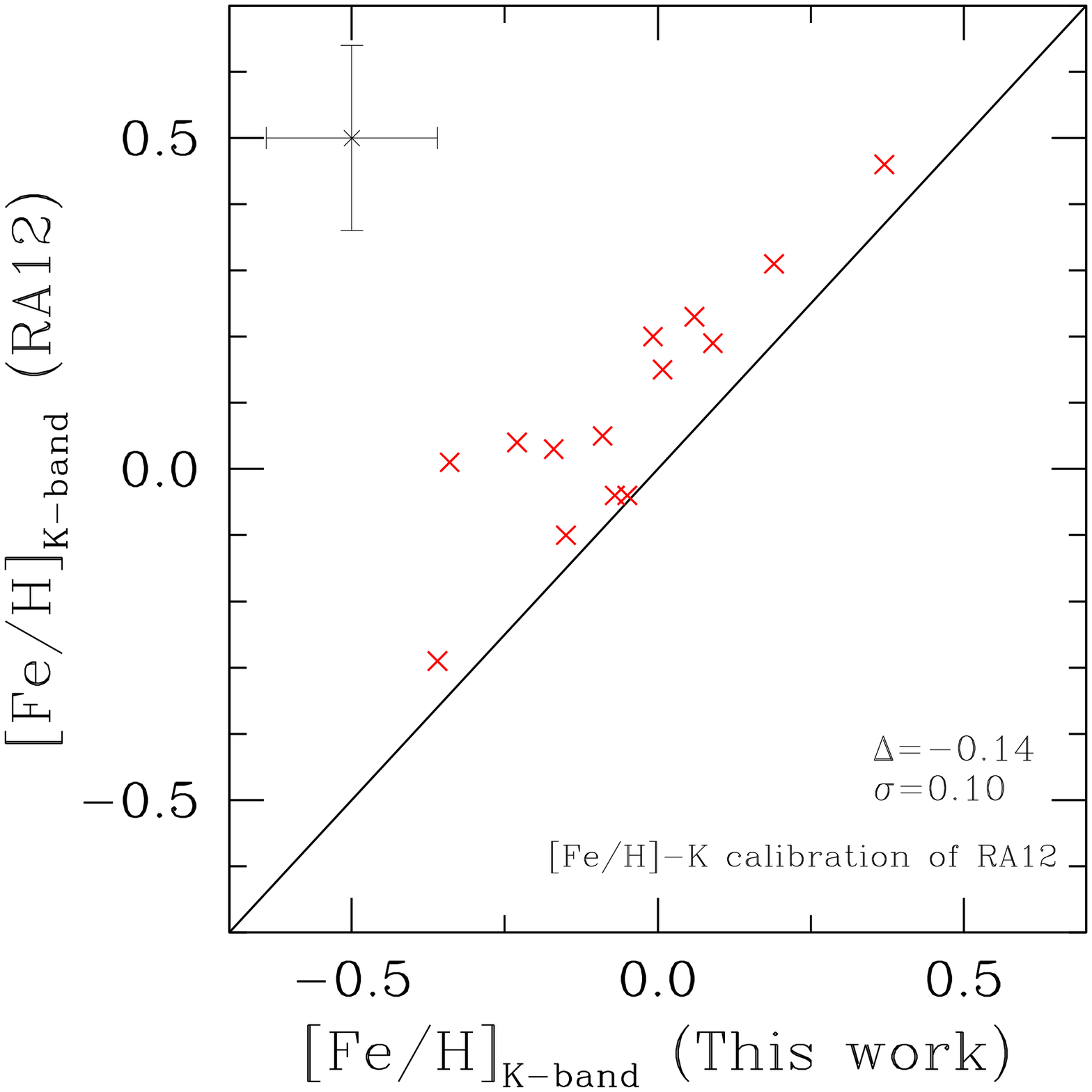}
	\includegraphics[width=0.3\columnwidth]{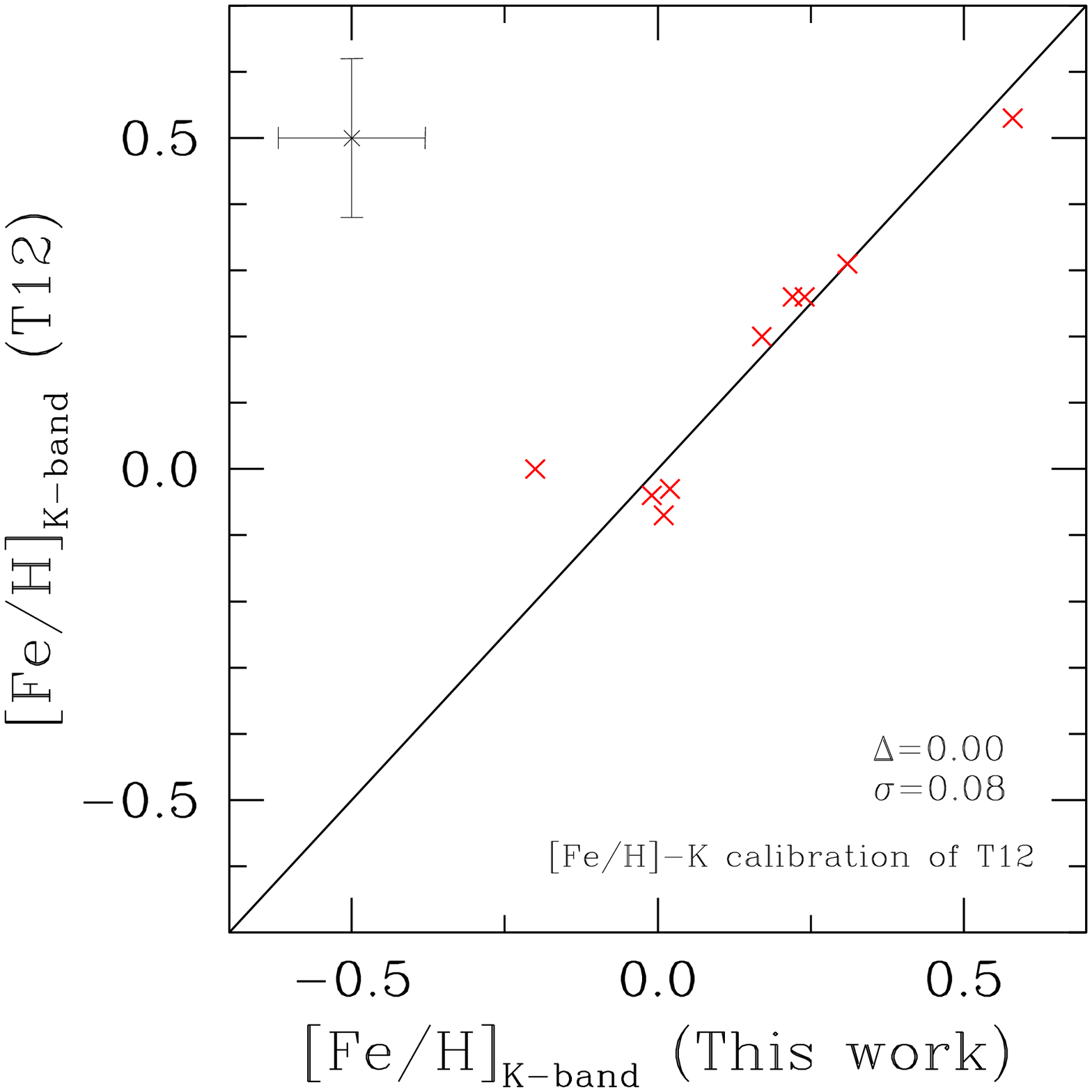}
	\includegraphics[width=0.3\columnwidth]{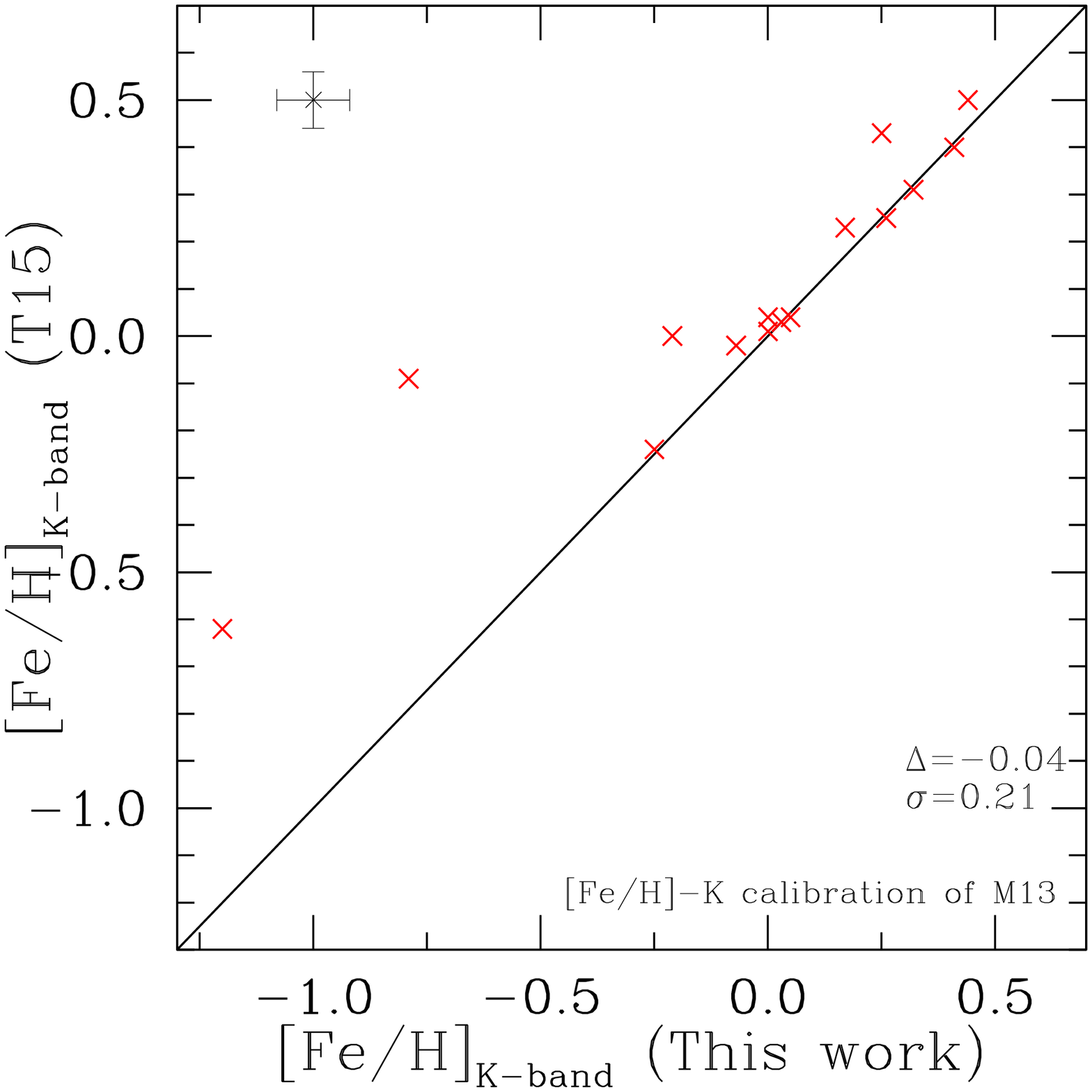}
	\caption{K-band calibrations. From left to right: Our calculated metallicities using \citet{RA12}'s calibrations
	vs those reported by these authors; metallicities derived using
	\citet{T12}'s calibrations vs those reported in that work; our
	determined metallicities applying \citet{Mann13}'s calibrations
	vs those reported in \citet{T15}. The black continuous lines correspond to the identity.
	In the case of the \citet{RA12}'s calibration our metallicities are
	systematically lower than those from the literature. On the other hand
	the comparison with \citet{T12} shows that our
	the metallicities are equivalent within error bars. With regard to \citet{T15},
	we find a good agreement except for two stars
	(GJ 250 B and GJ 611 B) with relatively low quality GNIRS spectra.}
	\label{comp_mets}
	\end{figure}
	\end{landscape}

	In order to check the consistency of our results with previous estimations, we compared
	the [Fe/H] of the stars we have in common with \citet{RA12}, \citet{T12}, and \citet{T15}.
	Unfortunately, \citet{Mann13} do not provide metallicities with their calibrations.
	Figure \ref{comp_mets} shows the comparison of our [Fe/H] values based on K-band calibrations to those from
	\citet{RA12}, \citet{T12}, and \citet{T15}, along with the median offset ($\Delta$\footnote{With $\Delta$ defined as
the median difference between our determinations and the literature values.}) and the standard
	deviation ($\sigma$) of the differences.
We found that our estimated [Fe/H] are, in general, smaller than the [Fe/H] listed by \citet{RA12}.
On the other hand, our measured [Fe/H] show very good agreement with those derived by \citet{T12}. In the case of the comparison with 
\citet{T15}, there are two stars (GJ 250 B and GJ 611 B)
for which the [Fe/H] we measure deviate considerably and systematically from the values obtained by 
these authors. As noted in section \ref{starmet} this may be attributed to the relatively low quality of our
spectra for these stars.

\begin{landscape}
	\begin{figure}
	\centering
	\includegraphics[width=0.3\columnwidth]{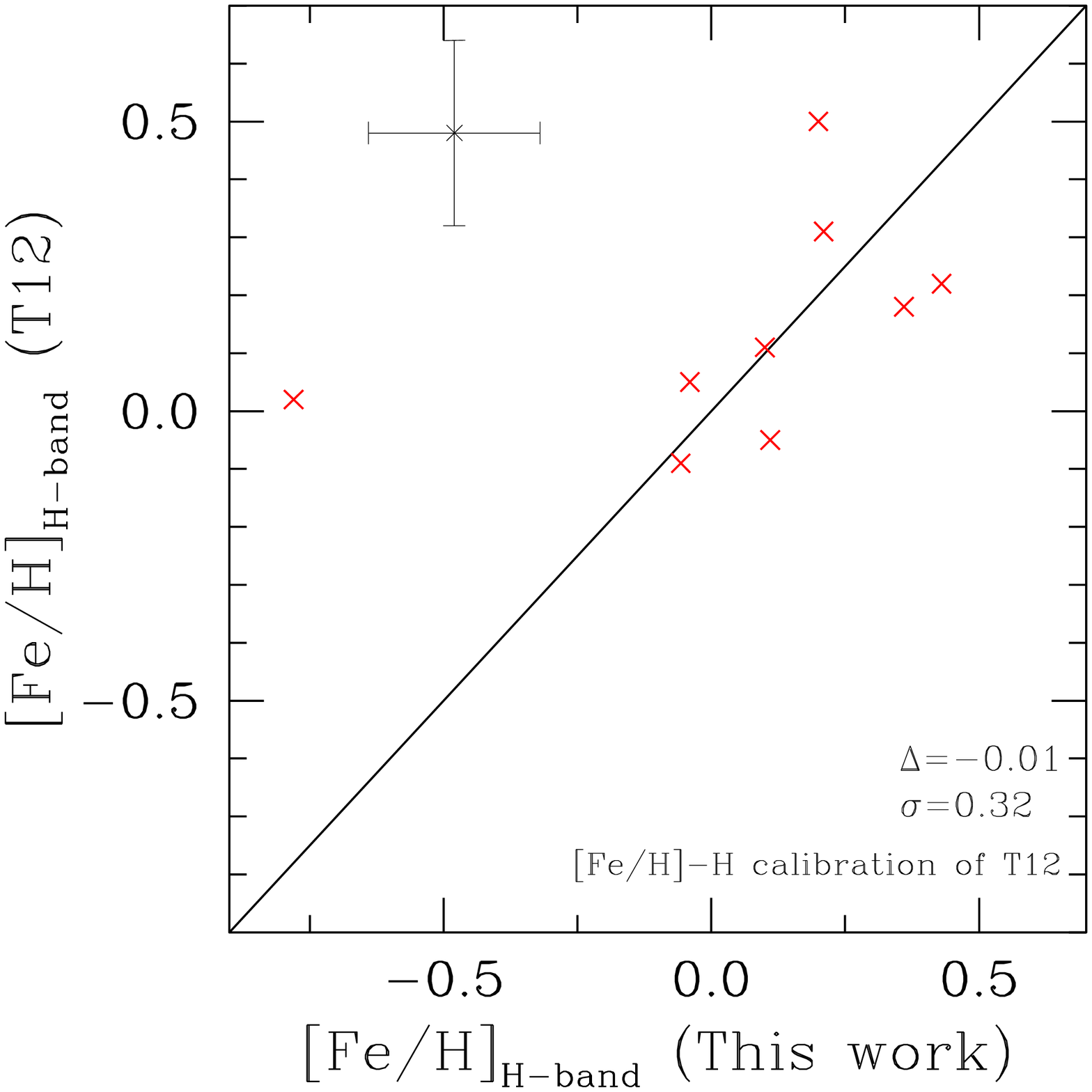}
	\includegraphics[width=0.3\columnwidth]{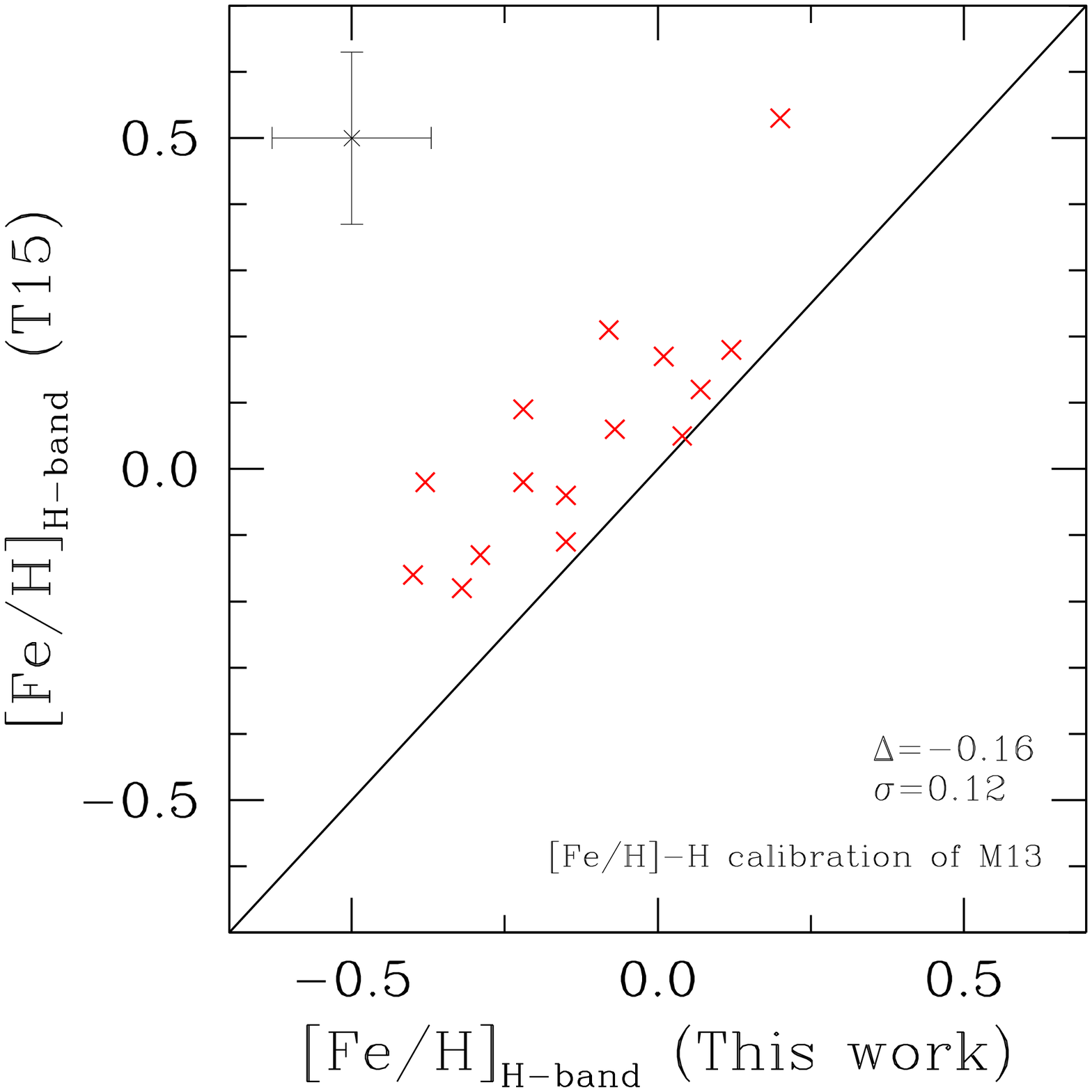}
	\caption{H-band calibrations. Left panel: Our calculated metallicities using \citet{T12}'s
	calibrations vs those reported by these authors. Right panel: 
	Metallicities using \citet{Mann13}'s calibrations vs those
	reported in \citet{T15}. The black continuous lines correspond to the identity.
        GNIRS measured [Fe/H] values are consistent with those of
        \citet{T12} within error bars, with exception of GJ 436, with 
        relatively low SNR spectra.  Our metallicities are in general $\sim$0.16 dex lower
        than those presented by \citet{T15}.}
	\label{comp_mets_band_h}
	\end{figure}
	\end{landscape}

	Analogously, Figure \ref{comp_mets_band_h} shows the comparison of our [Fe/H]
	values based on H-band calibrations
	with those from \citet{T12} and \citet{T15}. Our measured [Fe/H] values are consistent with those of
	\citet{T12} within error bars, with exception of GJ 436, for which our spectra has relatively low SNR 
        (see Table \ref{tab:Mstars}). 
	Our determinations are in general $\sim$0.16 dex lower than those presented
	by \citet{T15}. In both cases, as has been previously reported by \citet{Mann13}
	and \citet{T15}, it can be seen that the results using H-band calibrations show significantly larger scatter
	than the ones obtained from K-band calibrations.

	We also compared EWs and H$_2$O indices when possible. For \citet{RA12} we found that our EWs for Na and Ca are,
	in general, smaller for 13 stars. The average difference is of $\sim$10 \%. For \citet{T12} the comparison
	for 4 stars (GJ 250 B, GJ 297.2 B, GJ 777 B, and GJ 783.2 B) shows a better agreement with our measured
	EWs with an average
	difference of $\sim$5 \%. No equivalent widths are given in \citet{T15}.
	Comparisons of H$_2$O indices were possible only with those of \citet{RA12}. 
	In general, our calculated H$_2$O indices are larger than
	those reported by \citet{RA12} with an average difference of $\sim$8 \%. Larger water indices
        and smaller EWs are consistent
	with our [Fe/H] being smaller than those reported by \citet{RA12}.

	Although the spectral types of our stars fall within the range of
	the calibrations of \citet{RA12} and \citet{T12} (see section \ref{starmet})
	there is one star (HIP 79431) with [Fe/H] value outside the validity range of
	these two calibrations. In addition three others (GJ 179, GJ 436, and GJ 783.2 B) lie
	very close to the limits.
	Moreover, considering the results of \citet{Mann13} and 
	\citet{T15}, the good agreement between our [Fe/H]
	and those of \citet{T15} based on the K-band calibration of \citet{Mann13}, and that our
	spectra have similar resolution (R$\sim$1700) to those used to built
	the \citet{Mann13} calibrations, the following analyses are based exclusively
	on the K-band calibration of \citet{Mann13}.

	\section{Planet-metallicity correlation} \label{pmc}

	\subsection{[Fe/H] for M dwarfs with and without planets}

	As noted in \S~ \ref{intro}, the existence of a planet-metallicity correlation for M dwarfs has been
	strongly debated.  In this section, we use the metallicities obtained from GNIRS data and the 
	K-band calibration of \citet{Mann13} to analyze the [Fe/H] of M stars with a without planets. Then,
	using relatively large and homogeneous samples from \citet{T15}, we perform a similar analysis to verify that the
	trend found from our sample of 16 M-dwarf stars is consistent with the behaviour of larger samples.

\begin{center}
	\begin{figure}
	\centering
	\includegraphics[width=0.8\columnwidth]{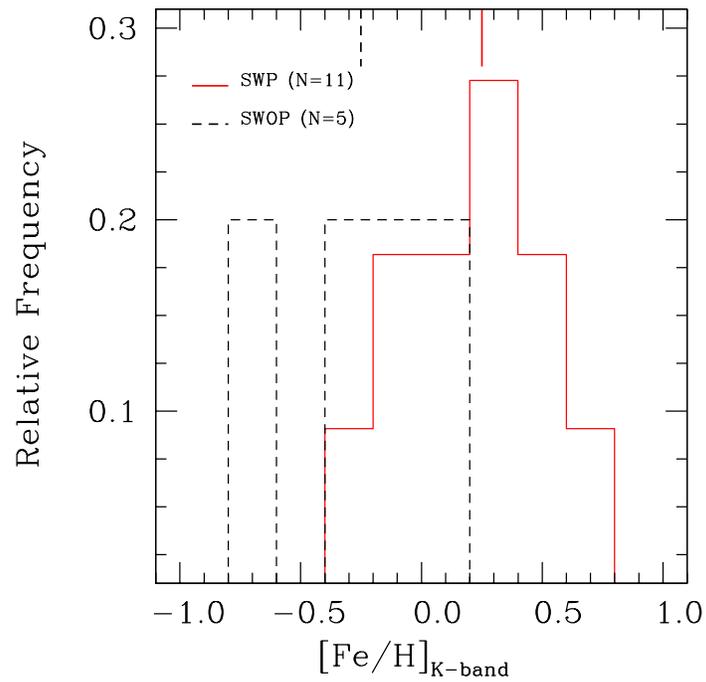}
	\caption{Metallicity distributions for stars with GNIRS spectra, derived using the calibration
	of \citet{Mann13} in the K band. Stars with planets (SWP, N$=$11) are indicated by the red continuous line
	and without planets (SWOP, N$=$5) by the black dash-line. The medians of both distributions, $+$0.25 and $-$0.25,
        are indicated.}
	\label{Hist-GNIRS-Mann13}
	\end{figure}
	\end{center}

	Figure \ref{Hist-GNIRS-Mann13} shows the [Fe/H] distributions of our sample of stars with planets
	(SWP, N$=$11, red continuous line) and without planets (SWOP, N$=$5, black dash-line).
	The corresponding medians are: 0.25 and $-$0.25, respectively. The distributions are
	significantly different, according to the Kolmogorov-Smirnov
	(K-S, p-value $=$ 0.05) test. 
	This result suggests that M-dwarfs with planets are, on average, more metallic than those without planets.
        Table \ref{stars-GNIRS} summarizes the statistics. 

        \begin{table}
        \centering
        \setlength{\tabnotewidth}{1\columnwidth}
        \tablecols{6}
        \caption{Metallicity distributions of the samples of M dwarfs with and without planet/s built
        from GNIRS spectra}
        \label{stars-GNIRS}
        \begin{tabular}{lccrcc}
        \toprule
        \\[-1em]
        Sample & Number & Average & Median & p-value (K-S)\tabnotemark{3} \\
        ID\tabnotemark{1}                 & of stars &  [Fe/H]\tabnotemark{2}& [Fe/H]\tabnotemark{2} & \\ \midrule
        SWP  & 11  & 0.21  & 0.25    & 0.01 \\
        SWOP  & 5  & $-$0.44  & $-$0.25 &   \\ \midrule

        \tabnotetext{1}{Notation: SWP: M dwarfs hosting planet/s;
        SWOP: M dwarfs without detected planet (see text for more details).}
        \tabnotetext{2}{Based on [Fe/H] values derived using GNIRS spectra and the K-band calibration of \citet{Mann13}.}
        \tabnotetext{3}{Probability of being drawn from the same distribution as the control sample, according to the
        two-sided Kolmogorov-Smirnov (K-S) test.}
        \end{tabular}
        \end{table}

	To test if the trend found from GNIRS spectra is supported by a larger sample,
	we used the catalogue of \citet{T15} that provides homogeneous [Fe/H] values for 886 M dwarfs, 
	based on the K-band calibration of \citet{Mann13}.  These authors flagged 16 planet hosts
	in their sample. We identified two additional stars with planets:
	GJ 176 \citep{Forveille2009} and GJ 3323 \citep{Astudillo-Defru2017}.
	The 18 M dwarfs with planets along with their [Fe/H] values are listed in
	Table \ref{Terrien-with-planets}.

	\begin{table}
	\centering
	\tablecols{3}
	\setlength{\tabnotewidth}{1\columnwidth}
	\caption{[Fe/H] of known M-dwarf-planet hosts from Terrien et al. (2015)'s catalogue}
	\label{Terrien-with-planets}
	\begin{tabular}{lrc}
	\toprule
	Name    &       [Fe/H]  &       Planet type\tabnotemark{$\dagger$}     \\ \bottomrule
	GJ 1214 &       0.40    &       super-Earth     \\
	GJ 176  &       0.23    &       super-Earth     \\
	GJ 179  &       0.25    &       Jupiter \\
	GJ 317  &       0.43    &       2 Jupiters      \\
	GJ 436  &       0.00    &       Neptune \\
	GJ 581  &       $-$0.02 &       Neptune $+$ 2 super-Earths      \\
	GJ 649  &       0.04    &       Jupiter $+$ super-Earth \\
	GJ 849  &       0.50    &       2 Jupiters      \\
	GJ 876  &       0.31    &       2 Jupiters $+$ Neptune $+$ super-Earth  \\
	HIP 57050       &       0.04    &       Jupiter \\
	HIP 79431       &       0.78    &       Jupiter \\
	WASP 43 &       0.40    &       Jupiter \\
	GJ 433  &       $-$0.03 &       Jupiter $+$ super-Earth \\
	WASP 80 &       0.13    &       Jupiter \\
	GJ 15 A &       $-$0.28 &       super-Earth     \\
	GJ 3323 &       $-$0.06 &       2 super-Earths  \\
	GJ 3470 &       0.27    &       Neptune \\
	Kepler 138      & $-$0.21       &       2 Earths $+$ Mars       \\ \bottomrule
	\tabnotetext{$\dagger$}{Classification derived from masses (and/or radii)
reported in The Extrasolar Planets Encyclopaedia (available at \url{www.exoplanet.eu}).}
	\end{tabular}
	\end{table}

	To construct a control sample of stars with no evidence of planetary companions,
	we cross-matched the remaining stars in \citet{T15} against samples of M dwarfs monitored
	with the HIRES  \citep{Rauscher-Marcy2006} and HARPS \citep{Bonfils2013} spectrographs, and
	selected all the stars without reported planets. In addition, we also
	included M dwarfs from the \citet{T15}'s catalogue with enough photometric measurements
	(usually more than $\sim$ 6500 data points) obtained with the \textit{Kepler} space mission
	and/or the SuperWASP ground-based survey and with no detected planets.
	In this way, we built a final control sample of 213 M dwarfs that were searched for planets,
	but for which no planet has been reported so far.  The selected M dwarfs without known planets along with
	the [Fe/H] values from \citet{T15}'s catalogue are listed in Table \ref{Terrien-without-planets}.
	We caution, however,  that this comparison sample might still include
	stars hosting low mass and/or long period planets that could be harder to detect by the mentioned surveys.

	{\tiny 
	\begin{center}
	\begin{longtable}{crcr}
	\caption{CONTROL SAMPLE WITHOUT KNOWN PLANETS AND K-BAND [Fe/H] FROM TERRIEN ET AL.'S CATALOGUE}\\
	\hline
	\label{Terrien-without-planets}

		Star	&	[Fe/H]	&		Star	&	[Fe/H]	\\ \bottomrule
	2MASSJ02361535$+$0652191	&	$-$0.29	&	2MASSJ12241121$+$2653166	&	$-$0.13	\\
	2MASSJ03132299$+$0446293	&	0.24	&	2MASSJ23465800$+$2750066	&	0.12	\\
	2MASSJ04425581$+$1857285	&	0.23	&	2MASSJ12214070$+$2707510	&	$-$0.1	0\\
	2MASSJ05015746$-$0656459	&	$-$0.06	&	2MASSJ23461405$+$2826036	&	$-$0.05	\\
	2MASSJ06521804$-$0511241	&	$-$0.09	&	2MASSJ00580115$+$3919111	&	$-$0.03	\\
	2MASSJ10121768$-$0344441	&	0.17	&	2MASSJ00270673$+$4941531	&	0.25	\\
	2MASSJ10285555$+$0050275	&	$-$0.19	&	2MASSJ09301445$+$2630250	&	0.22	\\
	2MASSJ10505201$+$0648292	&	0.29	&	2MASSJ02564122$+$3522346	&	0.24	\\
	2MASSJ11474440$+$0048164	&	$-$0.08	&	2MASSJ01270042$+$3351580	&	0.33	\\
	2MASSJ13295979$+$1022376	&	$-$0.12	&	2MASSJ13345147$+$3746195	&	0.21	\\
	2MASSJ14341683$-$1231106	&	0.42	&	2MASSJ04310001$+$3647548	&	0.12	\\
	2MASSJ15192689$-$0743200	&	$-$0.02	&	2MASSJ15512179$+$2931062	&	$-$0.11	\\
	2MASSJ17574849$+$0441405	&	$-$0.41	&	2MASSJ03302331$+$3440325	&	0.09	\\
	2MASSJ18050755$-$0301523	&	$-$0.26	&	2MASSJ21012481$+$2043377	&	$-$0.15	\\
	2MASSJ18424498$+$1354168	&	0.14	&	2MASSJ12362870$+$3512007	&	0.10	\\
	2MASSJ19095098$+$1740074	&	0.07	&	2MASSJ15493833$+$3448555	&	0.34	\\
	2MASSJ19220206$+$0702310	&	$-$0.28	&	2MASSJ02591670$+$3146245	&	0.01	\\
	2MASSJ20403364$+$1529572	&	0.33	&	2MASSJ20260528$+$5834224	&	0.30	\\
	2MASSJ22094029$-$0438267	&	0.50	&	2MASSJ12130291$+$2146388	&	$-$0.35	\\
	2MASSJ22531672$-$1415489	&	0.31	&	2MASSJ16061363$+$2901553	&	0.51	\\
	2MASSJ22563497$+$1633130	&	0.37	&	2MASSJ23435310$+$3235388	&	0.03	\\
	2MASSJ00182549$+$4401376	&	$-$0.1	&	2MASSJ17373648$+$2205510	&	0.08	\\
	2MASSJ02001278$+$1303112	&	$-$0.15	&	2MASSJ03360868$+$3118398	&	$-$0.04	\\
	2MASSJ02361535$+$0652191	&	$-$0.29	&	2MASSJ10361794$+$2844471	&	$-$0.18	\\
	2MASSJ06521804$-$0511241	&	$-$0.09	&	2MASSJ23442084$+$2136050	&	0.28	\\
	2MASSJ08524084$+$2818589	&	0.49	&	2MASSJ02000280$+$4345286	&	$-$0.09	\\
	2MASSJ10121768$-$0344441	&	0.17	&	2MASSJ10350859$+$3349499	&	$-$0.04	\\
	2MASSJ10285555$+$0050275	&	$-$0.19	&	2MASSJ02000280$+$4345286	&	$-$0.09	\\
	2MASSJ10505201$+$0648292	&	0.29	&	2MASSJ23385568$+$2101218	&	0.11	\\
	2MASSJ11474440$+$0048164	&	$-$0.08	&	2MASSJ23071524$-$2307533	&	$-$0.06	\\
	2MASSJ13295979$+$1022376	&	$-$0.12	&	2MASSJ22225080$+$2801475	&	0.22	\\
	2MASSJ14341683$-$1231106	&	0.42	&	2MASSJ02132062$+$3648506	&	$-$0.05	\\
	2MASSJ17574849$+$0441405	&	$-$0.41	&	2MASSJ15383708$+$3707247	&	0.01	\\
	2MASSJ18050755$-$0301523	&	$-$0.26	&	2MASSJ11263757$+$3756237	&	0.15	\\
	2MASSJ18343664$+$4007266	&	0.84	&	2MASSJ03564330$+$3254082	&	0.07	\\
	2MASSJ22464980$+$4420030	&	0.05	&	2MASSJ18562628$+$4622532	&	$-$0.01	\\
	2MASSJ22563497$+$1633130	&	0.37	&	2MASSJ03323578$+$2843554	&	$-$0.19	\\
	2MASSJ23415498$+$4410407	&	0.55	&	2MASSJ13455527$+$2723131	&	0.18	\\
	2MASSJ18550451$+$4259510	&	0.11	&	2MASSJ03323578$+$2843554	&	$-$0.19	\\
	2MASSJ19051335$+$3845050	&	0.06	&	2MASSJ23450868$+$3003184	&	$-$0.15	\\
	2MASSJ19051739$+$4507161	&	$-$0.19	&	2MASSJ18523373$+$4538317	&	$-$0.04	\\
	2MASSJ19170558$+$4007235	&	$-$0.19	&	2MASSJ11281625$+$3136017	&	0.20	\\
	2MASSJ19242100$+$4237254	&	0.44	&	2MASSJ13514938$+$4157445	&	0.39	\\
	2MASSJ19271753$+$4231537	&	$-$0.07	&	2MASSJ13505181$+$3644168	&	$-$0.07	\\
	2MASSJ19510930$+$4628598	&	$-$0.05	&	2MASSJ11353198$+$3855372	&	0.08	\\
	2MASSJ17283039$+$3727074	&	$-$0.09	&	2MASSJ12242665$+$2545077	&	$-$0.06	\\
	2MASSJ17074035$+$4918351	&	0.01	&	2MASSJ16041322$+$2331386	&	0.13	\\
	2MASSJ17340562$+$4447082	&	0.23	&	2MASSJ11240434$+$3808108	&	0.00	\\
	2MASSJ17363485$+$4549324	&	0.40	&	2MASSJ02224082$+$3055161	&	$-$0.03	\\
	2MASSJ16352740$+$3500577	&	$-$0.06	&	2MASSJ10331367$+$3409120	&	$-$0.15	\\
	2MASSJ17302672$+$3344522	&	0.26	&	2MASSJ22182135$+$4356406	&	$-$0.15	\\
	2MASSJ17173857$+$5224227	&	$-$0.01	&	2MASSJ01401649$+$3147306	&	0.25	\\
	2MASSJ17032384$+$5124219	&	0.03	&	2MASSJ10335971$+$2922465	&	0.03	\\
	2MASSJ16454410$+$3605496	&	$-$0.22	&	2MASSJ00252063$+$2253121	&	$-$0.01	\\
	2MASSJ17092601$+$3909384	&	$-$0.2	&	2MASSJ00243478$+$3002295	&	0.23	\\
	2MASSJ17080710$+$4829268	&	$-$0.02	&	2MASSJ21395433$+$2736439	&	$-$0.25	\\
	2MASSJ17072670$+$3900429	&	0.11	&	2MASSJ20592035$+$5303049	&	0.13	\\
	2MASSJ17101101$+$4139340	&	0.11	&	2MASSJ17195948$+$2412054	&	0.10	\\
	2MASSJ23225835$+$3717143	&	$-$0.10	&	2MASSJ17190577$+$2253036	&	0.35	\\
	2MASSJ17101101$+$4139340	&	0.11	&	2MASSJ00285391$+$5022330	&	0.15	\\
	2MASSJ22294885$+$4128479	&	$-$0.02	&	2MASSJ13093495$+$2859065	&	0.02	\\
	2MASSJ16495034$+$4745402	&	0.16	&	2MASSJ05030563$+$2122362	&	0.07	\\
	2MASSJ16480454$+$4522429	&	0.09	&	2MASSJ12462672$+$2626368	&	0.08	\\
	2MASSJ15315427$+$2851096	&	0.19	&	2MASSJ12503457$+$2655230	&	$-$0.1	\\
	2MASSJ16533915$+$5603272	&	$-$0.22	&	2MASSJ04342248$+$4302148	&	0.22	\\
	2MASSJ16312806$+$4710212	&	0.04	&	2MASSJ13220965$+$4144432	&	$-$0.12	\\
	2MASSJ02000741$+$3639481	&	0.10	&	2MASSJ14170294$+$3142472	&	0.10	\\
	2MASSJ12265737$+$2700536	&	0.11	&	2MASSJ07003840$+$3334581	&	0.16	\\
	2MASSJ23292258$+$4127522	&	0.24	&	2MASSJ00383388$+$5127579	&	$-$0.17	\\
	2MASSJ16342040$+$5709439	&	$-$0.4	&	2MASSJ16505794$+$2227058	&	$-$0.16	\\
	2MASSJ12255421$+$2651387	&	$-$0.07	&	2MASSJ00383388$+$5127579	&	$-$0.17	\\
	2MASSJ22011310$+$2818248	&	0.02	&	2MASSJ03563308$+$3157248	&	0.21	\\
	2MASSJ18352722$+$4545403	&	$-$0.17	&	2MASSJ16541912$+$2537363	&	$-$0.14	\\
	2MASSJ16495777$+$4601418	&	0.10	&	2MASSJ04040615$+$3042454	&	$-$0.39	\\
	2MASSJ23384176$+$3909262	&	0.08	&	2MASSJ16541912$+$2537363	&	$-$0.14	\\
	2MASSJ17555802$+$2926097	&	0.15	&	2MASSJ23215594$+$2412321	&	0.12	\\
	2MASSJ22172586$+$2335047	&	0.04	&	2MASSJ23495384$+$2721406	&	0.02	\\
	2MASSJ12250262$+$2642382	&	0.04	&	2MASSJ19071270$+$4416070	&	0.31	\\
	2MASSJ17393223$+$2746366	&	0.08	&	2MASSJ22384426$+$2513305	&	0.08	\\
	2MASSJ13314666$+$2916368	&	0.12	&	2MASSJ08175130$+$3107455	&	0.27	\\
	2MASSJ11315396$+$2725336	&	0.73	&	2MASSJ01382392$+$4516549	&	$-$0.24	\\
	2MASSJ13323908$+$3059065	&	0.17	&	2MASSJ00115302$+$2259047	&	0.25	\\
	2MASSJ18180345$+$3846359	&	$-$0.15	&	2MASSJ01040580$+$3938159	&	0.06	\\
	2MASSJ22232904$+$3227334	&	0.12	&	2MASSJ15294392$+$4252498	&	0.03	\\
	2MASSJ02591060$+$3636402	&	$-$0.01	&	2MASSJ05295269$+$3204524	&	$-$0.25	\\
	2MASSJ23575452$+$2159281	&	0.23	&	2MASSJ12232063$+$2529441	&	0.15	\\
	2MASSJ16071362$+$2650173	&	$-$0.2	&	2MASSJ09370355$+$4034389	&	0.15	\\
	2MASSJ13332256$+$3620352	&	0.37	&	2MASSJ23454076$+$4942300	&	0.23	\\
	2MASSJ17002033$+$2521028	&	$-$0.18	&	2MASSJ08155393$+$3136392	&	0.28	\\
	2MASSJ00085391$+$2050252	&	0.15	&	2MASSJ09093060$+$3249091	&	$-$0.22	\\
	2MASSJ12305549$+$3152121	&	0.16	&	2MASSJ21362954$+$5331585	&	0.14	\\
	2MASSJ13451104$+$2852012	&	$-$0.11	&	2MASSJ12424996$+$4153469	&	0.30	\\
	2MASSJ23295502$+$2211442	&	$-$0.36	&	2MASSJ23454076$+$4942300	&	0.23	\\
	2MASSJ12292712$+$2259467	&	0.00	&	2MASSJ21362954$+$5331585	&	0.14	\\
	2MASSJ21395433$+$2736439	&	$-$0.25	&	2MASSJ23454076$+$4942300	&	0.23	\\
	2MASSJ21415843$+$2741150	&	$-$0.20	&	2MASSJ10145315$+$2123464	&	0.07	\\
	2MASSJ21395433$+$2736439	&	$-$0.25	&	2MASSJ01512417$+$2123399	&	0.17	\\
	2MASSJ21415843$+$2741150	&	$-$0.2	&	2MASSJ21274751$+$5505337	&	$-$0.21	\\
	2MASSJ01031395$+$3140598	&	0.30	&	2MASSJ21462206$+$3813047	&	$-$0.56	\\
	2MASSJ14412571$+$2839269	&	0.28	&	2MASSJ19562490$+$5909216	&	$-$0.44	\\
	2MASSJ23422211$+$3458276	&	0.12	&	2MASSJ02085359$+$4926565	&	0.13	\\
	2MASSJ23545147$+$3831363	&	0.14	&	2MASSJ06222070$+$3326564	&	0.39	\\
	2MASSJ12573935$+$3513194	&	$-$0.08	&	2MASSJ10494561$+$3532515	&	$-$0.49	\\
	2MASSJ16043696$+$2620430	&	$-$0.01	&	2MASSJ21462206$+$3813047	&	$-$0.56	\\
	2MASSJ23425274$+$3049219	&	0.05	&	2MASSJ23565510$+$2305033	&	$-$0.07	\\
	2MASSJ23423350$+$3914234	&	$-$0.39	&	2MASSJ01431186$+$2101106	&	0.33	\\
	2MASSJ23505402$+$3829334	&	0.39	&	         			&               \\
	\hline
	\end{longtable}
	\end{center}
	}

\begin{landscape}
	\begin{figure}
	\centering
	\includegraphics[width=0.4\columnwidth]{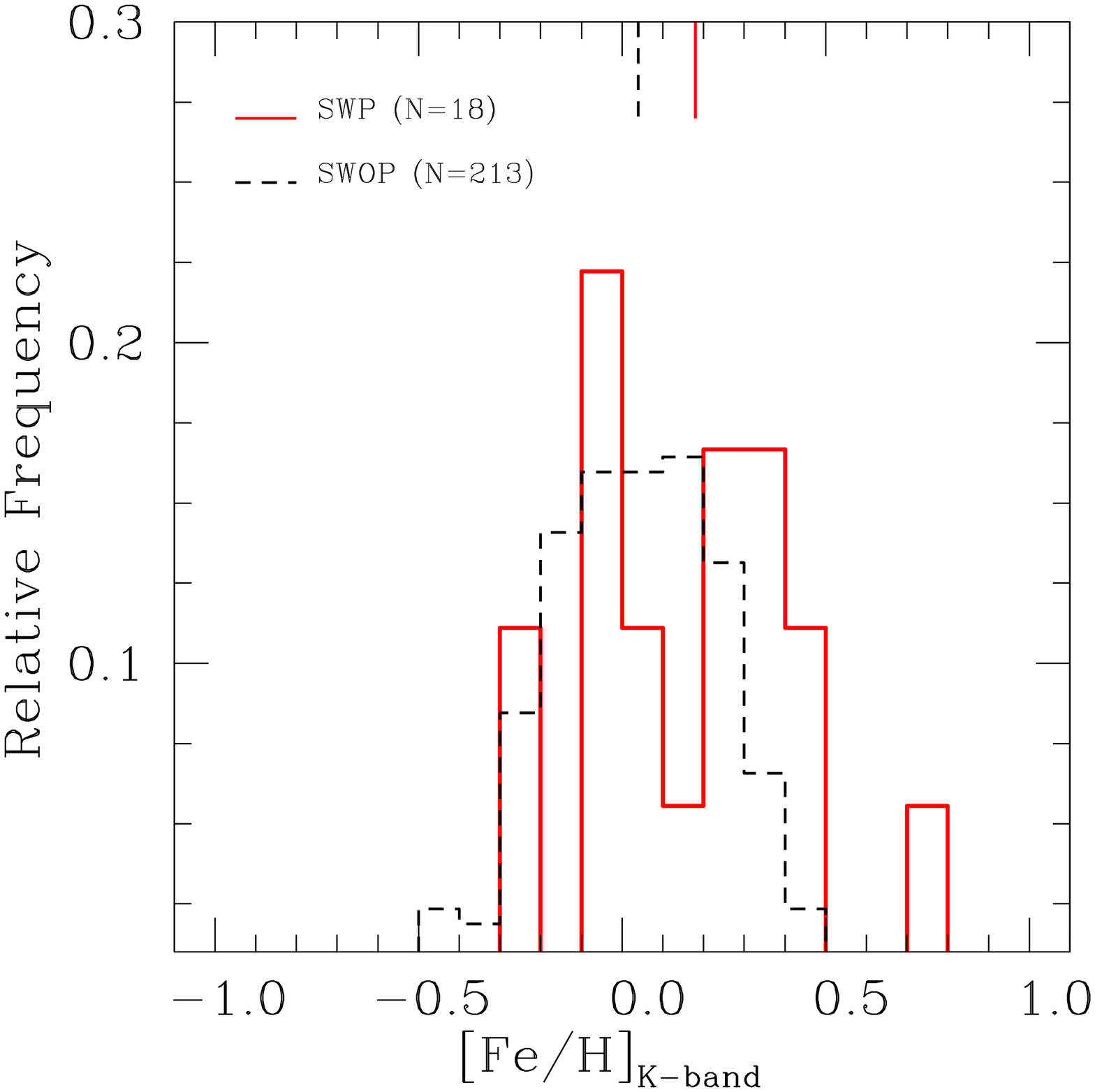}
	\includegraphics[width=0.4\columnwidth]{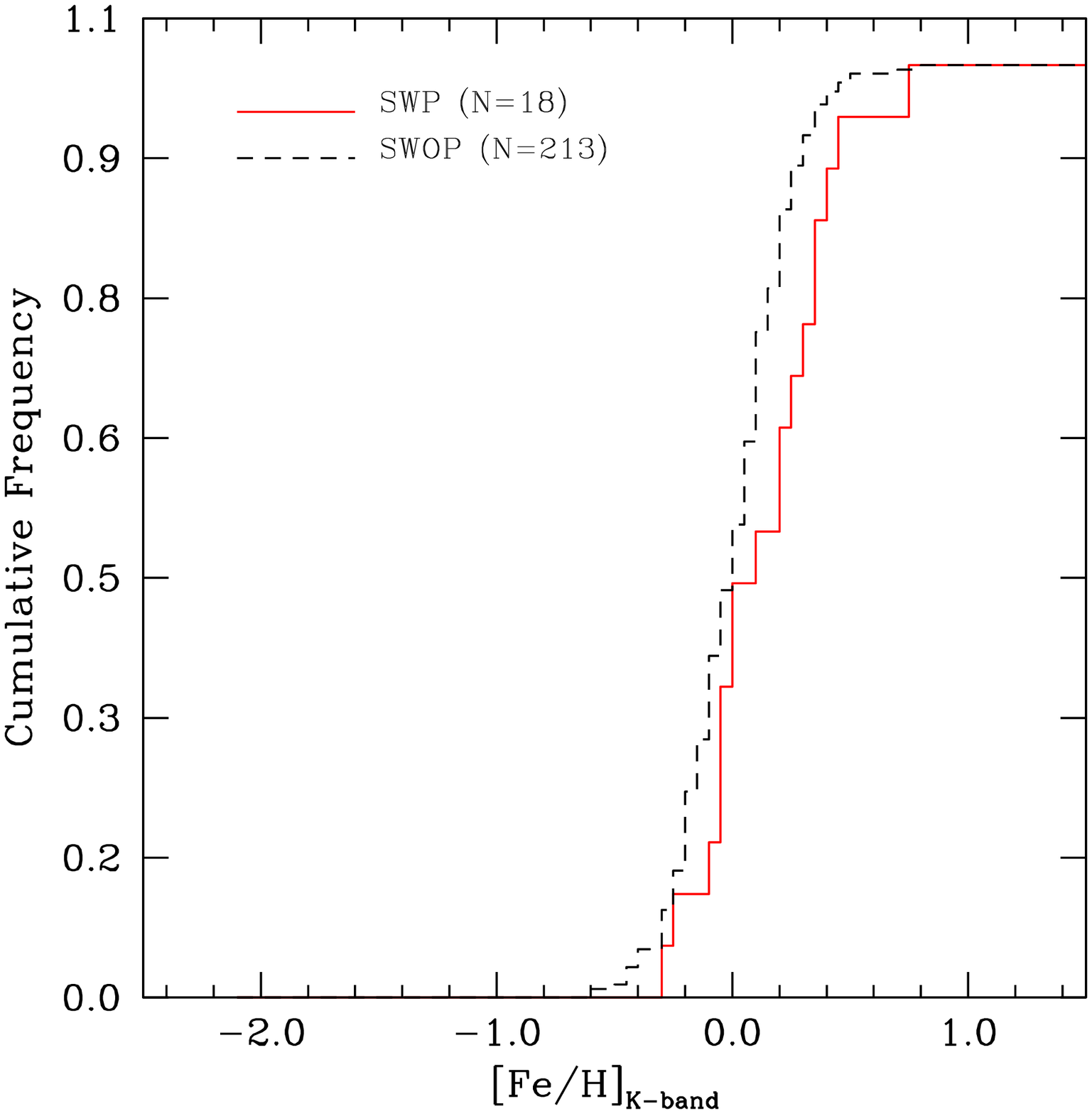}
	\caption{Normalised metallicity distributions and cumulative frequencies
	for the M dwarfs with planets (SWP, N$=$18, red continuous line) and without known planets
	(SWOP, N$=$213, black dashed line), based on K-band [Fe/H] values from \citet{T15}.}
	\label{stars-from-T15}
	\end{figure}
	\end{landscape}

	Figure \ref{stars-from-T15} shows the normalised metallicity histograms along with the
	cumulative frequencies for the M dwarfs with planets (SWP, N$=$18, red continuous line)
	and without known planets (SWOP, N$=$213, black dashed line), based on K-band [Fe/H] values from \citet{T15}. 
	The sample of M dwarfs with planets has a median metallicity of $+$0.18 dex, whereas 
	the control sample has a median of 0.05 dex. The two-sided K-S test gives a probability of 0.11
	that both samples share the same parent distribution. 
	Table \ref{summary} summarizes the statistics.
	The distribution of M dwarfs with planets is shifted toward higher metallicities with respect to
	the control sample by $\sim$ 0.11 dex, showing the same trend found from 
	the sample based on GNIRS data. 
	This result is in agreement with the planet-metalliticy correlation for M dwarfs with planets
	found by other authors (e.g. \citealt{JohnsonApps},  \citealt{RA10},
	\citealt{T12}, \citealt{Neves}, \citealt{GaidosMann})
	and it is also consistent with the metallicity
	enhancement found in solar-type stars with planets
	(e.g. \citealt{FischerValenti}, \citealt{Santos2004}, \citealt{Santos2005},
	\citealt{GhezziA}, \citealt{Maldonado}). 

        \begin{table}
        \centering
        \setlength{\tabnotewidth}{1\columnwidth}
        \tablecols{6}
        \caption{Metallicity distributions of M dwarfs with planet/s and the control sample built
        from TERRIEN ET AL.'S catalogue}
        \label{summary}
        \begin{tabular}{lccrcc}
        \toprule
        \\[-1em]
        Sample & Number & Average & Median & p-value (K-S)\tabnotemark{3} \\
        ID\tabnotemark{1}                 & of stars &  [Fe/H]\tabnotemark{2}& [Fe/H]\tabnotemark{2} & \\ \midrule
        SWP  & 18  & 0.18  & 0.18 & 0.11 \\
        SWOP  & 213  & 0.04  & 0.05 &  \\ \midrule

        \tabnotetext{1}{Notation: SWP: M dwarfs hosting planet/s;
        SWOP: control sample without detected planets (see text for more details).}
        \tabnotetext{2}{Based on [Fe/H] values derived by \citet{T15} using the K-band calibration of \citet{Mann13}.}
        \tabnotetext{3}{Probability of being drawn from the same distribution as the control sample, according to the
        two-sided Kolmogorov-Smirnov (K-S) test.}
        \end{tabular}
        \end{table}

	\subsection{Metallicity vs planetary parameters} \label{mpp}

	To investigate whether correlations between [Fe/H] and planetary parameters
	found for FGK main-sequence stars (e.g. \citealt{FischerValenti}, \citealt{Bouchaave2012},
	\citealt{Adib}) 
	are valid in M-type stars we searched for correlations of
	[Fe/H] with the planetary masses o planet types, orbital periods, and eccentricities. For this analysis,
	we first used the 11 stars with planets observed with GNIRS and then 
	the 18 stars with planets (which also includes the stars with planets observed with GNIRS)
	and the control sample derived from \citet{T15}'s catalogue (see Table \ref{Terrien-without-planets}). 
	Planetary parameters (masses, periods, and eccentricities) were
	obtained from The Extrasolar Planets Encyclopaedia.

	\subsubsection{Metallicity vs M-dwarfs hosting Jovian-mass and lower-mass planets}

	We constructed two sub-samples from the 11 M-dwarf stars with planets (SWP) observed with GNIRS:
	stars with at least one Jupiter-mass planet (SWJP) and those hosting only lower mass planets ($M_{p}\sin i$
	$<$ 25 $M_{\oplus}$, mainly Neptune- and super-Earth mass planets, SWLP). In a similar manner,
	we classified the sample of 18 M-dwarf stars with planets from \citet{T15} in two sub-sets. 

\begin{landscape}
\begin{figure}
\centering
\includegraphics[width=0.4\columnwidth]{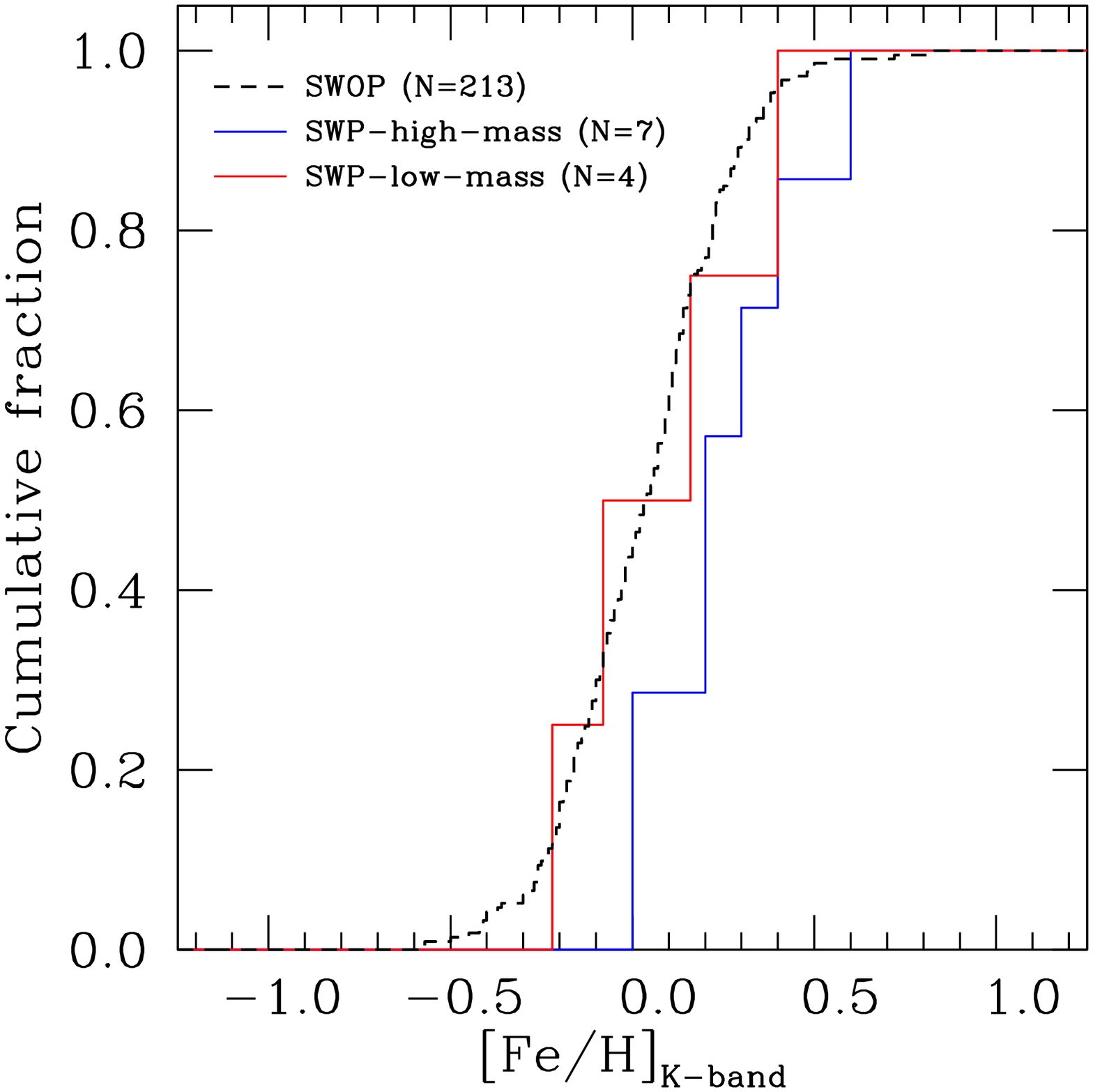}
\includegraphics[width=0.4\columnwidth]{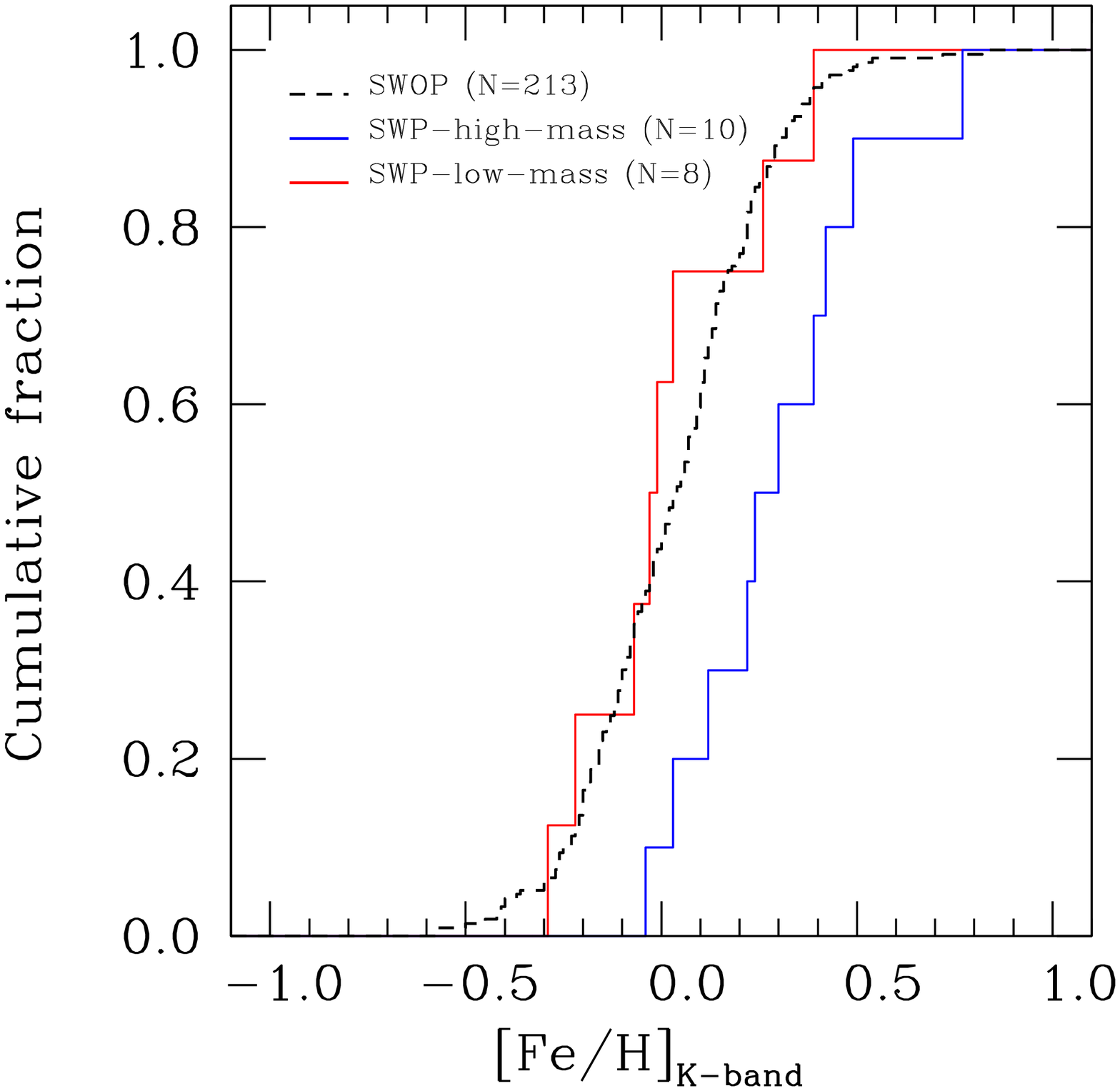}
\caption{Cumulative frequency distributions of M dwarfs hosting high-mass planets
(SWP-high-mass blue continuous line) and low-mass planets (SWP-low-mass red continuous line),
in addition to the control sample composed by stars without known planets (SWOP black dashed line) taken
from \citet{T15}'s catalogue (see Table \ref{Terrien-without-planets}).
The left panel shows only the metallicties derived from GNIRS spectra, except for the control sample.
The right panel shows metallicities from \citet{T15}.}
\label{stars-mass-from-T15}
\end{figure}
\end{landscape}

	Figure \ref{stars-mass-from-T15} shows the cumulative frequency distributions for Jupiter-like hosts
	(blue continuous line), low-mass
	(Neptune and super-Earth type) planets (red continuous line) in comparison with the control sample
	(black dashed line) from \citet{T15}'s catalogue.  The left panel corresponds to the GNIRS sample,
	the right panel to the \citet{T15} sub-set. Table \ref{summary1-Terrien} summarizes the statistics for
	both samples. We note that both samples of stars with planets are
	compared with the \citet{T15} control sample.  We observed only 5 stars without planets with GNIRS
	(see Table \ref{tab:Mstars}). However the trend in Table \ref{summary1-Terrien} for GNIRS stars with
	planets remains even if such a small control sample is used. 

	The K-S test gives p-values of 0.02 and 0.01 (GNIRS and \citet{T15}'s samples, respectively) that
	the hosts of giant planets and the M-dwarf control
	sample are drawn from the same parent distribution.  The metallicity distributions
	of Jupiter-like hosts are clearly shifted, by $\sim$ $+$0.20 dex, to higher metallicities compared
	with the control sample, for both GNIRS and \citet{T15}'s samples. On the other hand, the [Fe/H]
	distributions of stars hosting low-mass planets and those without known planets are very similar. 
	In this case, the K-S test gives p-values of 0.97 and 0.75 (GNIRS and \citet{T15}'s samples) that
	Neptune and super-Earth hosts share the same metallicity distribution as the control sample. 

	It must be noted, however, that the orbital inclinations of the planets  - and hence the true masses -
	have only been determined for a third of the planets under consideration. For the remaining planets,
	only the lower bound of $M sin(i)$ is known.
	In addition we caution about the small number of objects in both
	sub-samples with planets and the need to increase the number of M dwarfs with planets to
	put this initial result on more solid grounds. On the other hand, we note that in spite of the
	small numbers of M-dwarfs with planets, GNIRS based samples agree with \citet{T15}'s.

	This analysis suggests that, like their more massive solar-type counterparts
	(\citealt{Sousa2008}, \citealt{GhezziA}, \citealt{Mayor}, \citealt{Bouchaave2012},
	\citealt{Neves}), M dwarfs hosting low-mass planets are not preferentially metal-rich.
	In addition, this result is in line with similar suggestions obtained from smaller samples of M dwarfs
	with planets (\citealt{JohnsonApps}, \citealt{RA12}, \citealt{T12}, \citealt{GaidosMann}).

	The apparent separation in metallicity between host stars harboring only Neptune and/or
	super-Earth type planets and those with at least one Jupiter-type planet can be explained in the context
	of the core accretion model of planetary formation. This model postulates that only metal-rich disks would
	form cores rapidly enough to allow for gas accretion on a sufficient scale
	as to form Jupiter-type planets before the gas dissipates, as described in \S~\ref{intro}.

\begin{table}
\centering
\setlength{\tabnotewidth}{1\columnwidth}
\tablecols{6}
\caption{Metallicity distributions of different M-dwarf samples with giant and low-mass planets}
\label{summary1-Terrien}
\begin{tabular}{lccrcc}
\toprule
\\[-1em]
Sample & Number & Average & Median & p-value (K-S)\tabnotemark{3} \\
ID\tabnotemark{1}                 & of stars &  [Fe/H]\tabnotemark{2}& [Fe/H]\tabnotemark{2} & \\ \midrule
\multicolumn{4}{c}{GNIRS sample}\\ \midrule
SWJP  & 7  & 0.28  & 0.26 & 0.02 \\
SWLP  & 4 & 0.07  & 0.05 & 0.97 \\ \midrule
\multicolumn{4}{c}{\citet{T15}'s catalogue}\\ \midrule
SWJP  & 10  & 0.30  & 0.28 & 0.01 \\
SWLP  & 8 & 0.02  & $-$0.01 & 0.75 \\ \midrule

\tabnotetext{1}{Notation: SWJP: M dwarfs harbouring at least one giant planet;
SWLP: M dwarfs hosting only low-mass planets.}
\tabnotetext{2}{Based on [Fe/H] values derived using the K-band calibration of \citet{Mann13}.}
\tabnotetext{3}{Probability of being drawn from the same distribution as the control sample, according to the
two-sided Kolmogorov-Smirnov (K-S) test.}
\end{tabular}
\end{table}

\subsubsection{Metallicity vs planetary periods and eccentricities}

\begin{landscape}
\begin{figure}
\begin{center}
\includegraphics[width=0.34\columnwidth]{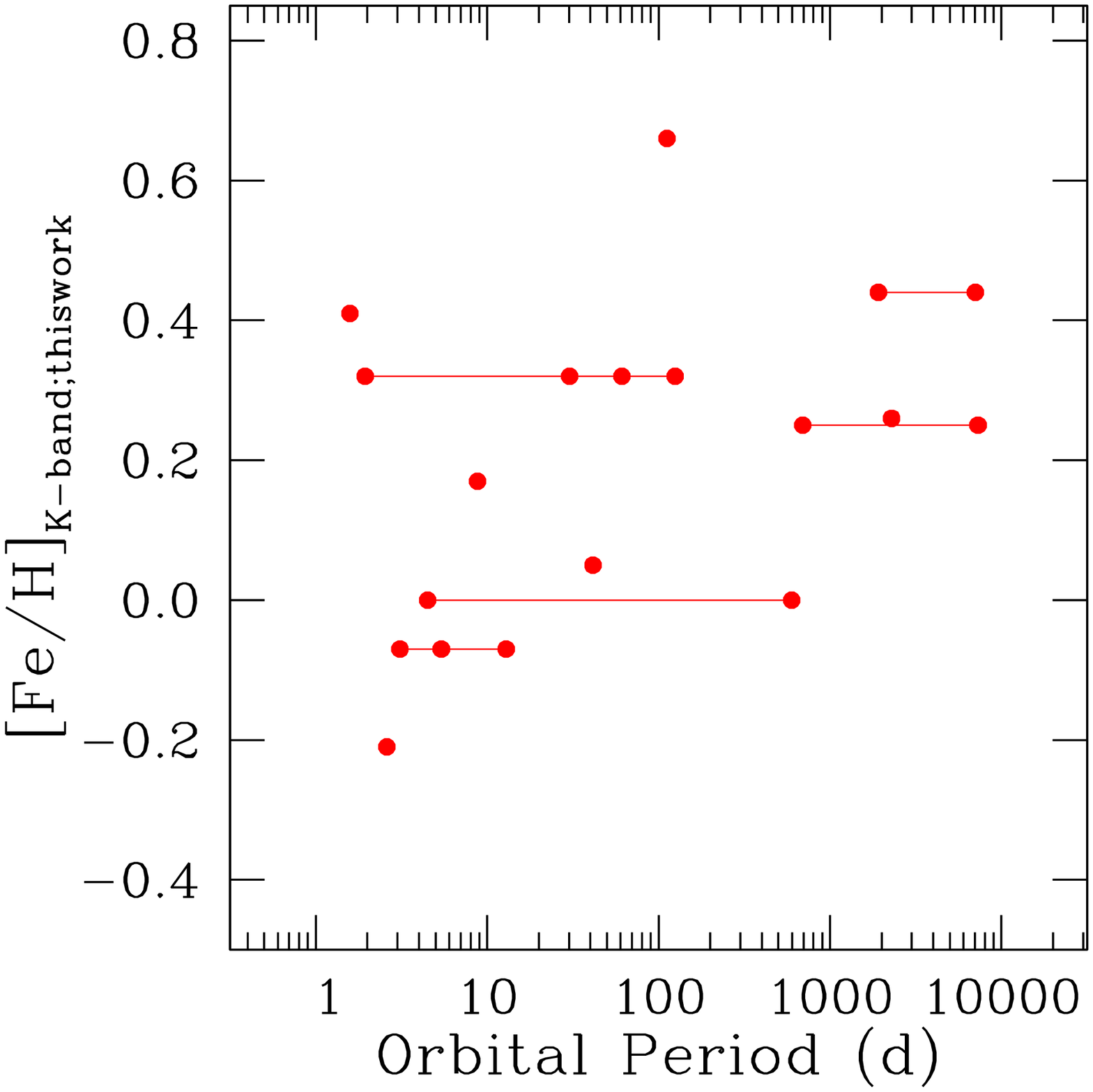}
\includegraphics[width=0.34\columnwidth]{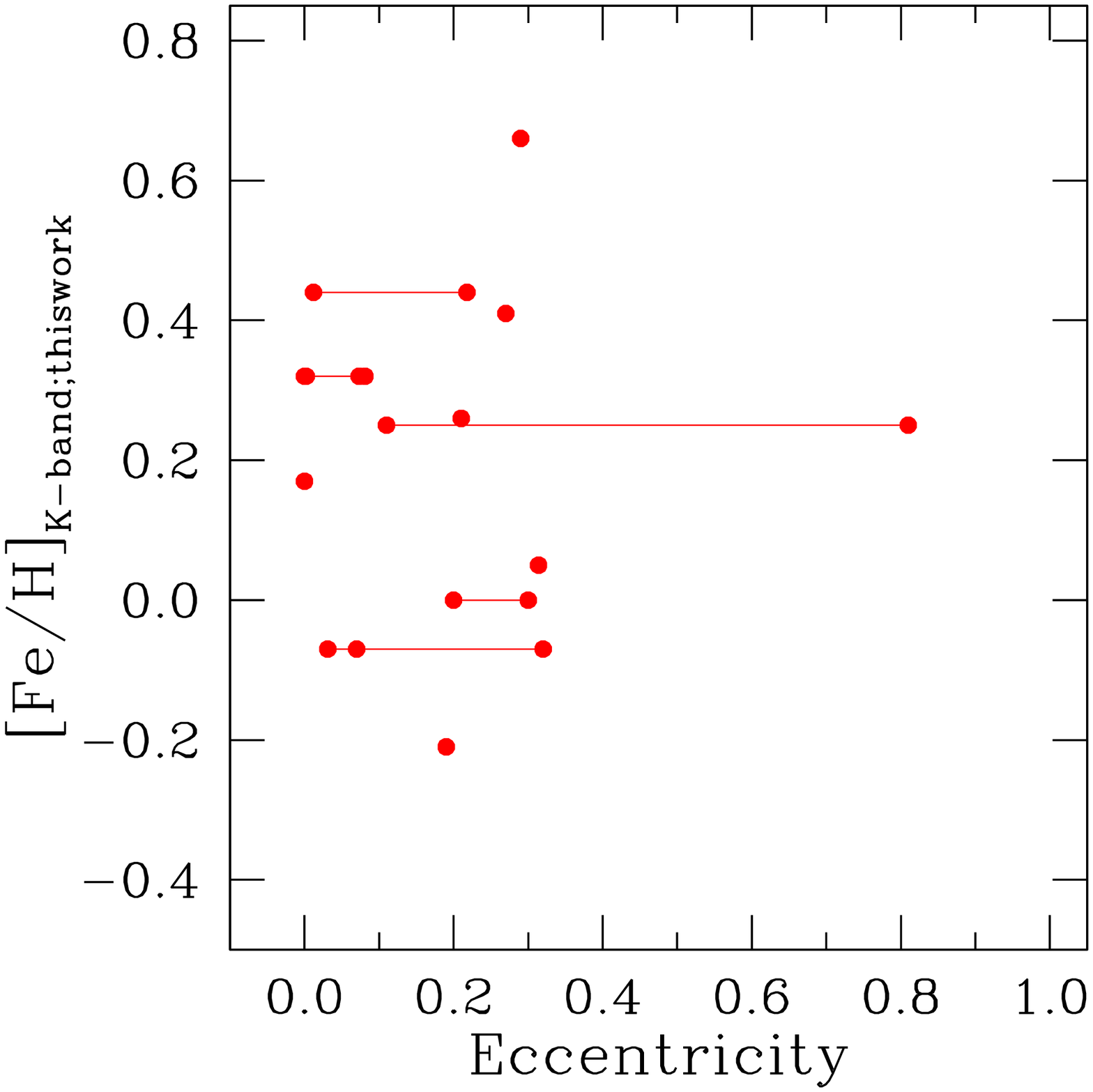}
\includegraphics[width=0.34\columnwidth]{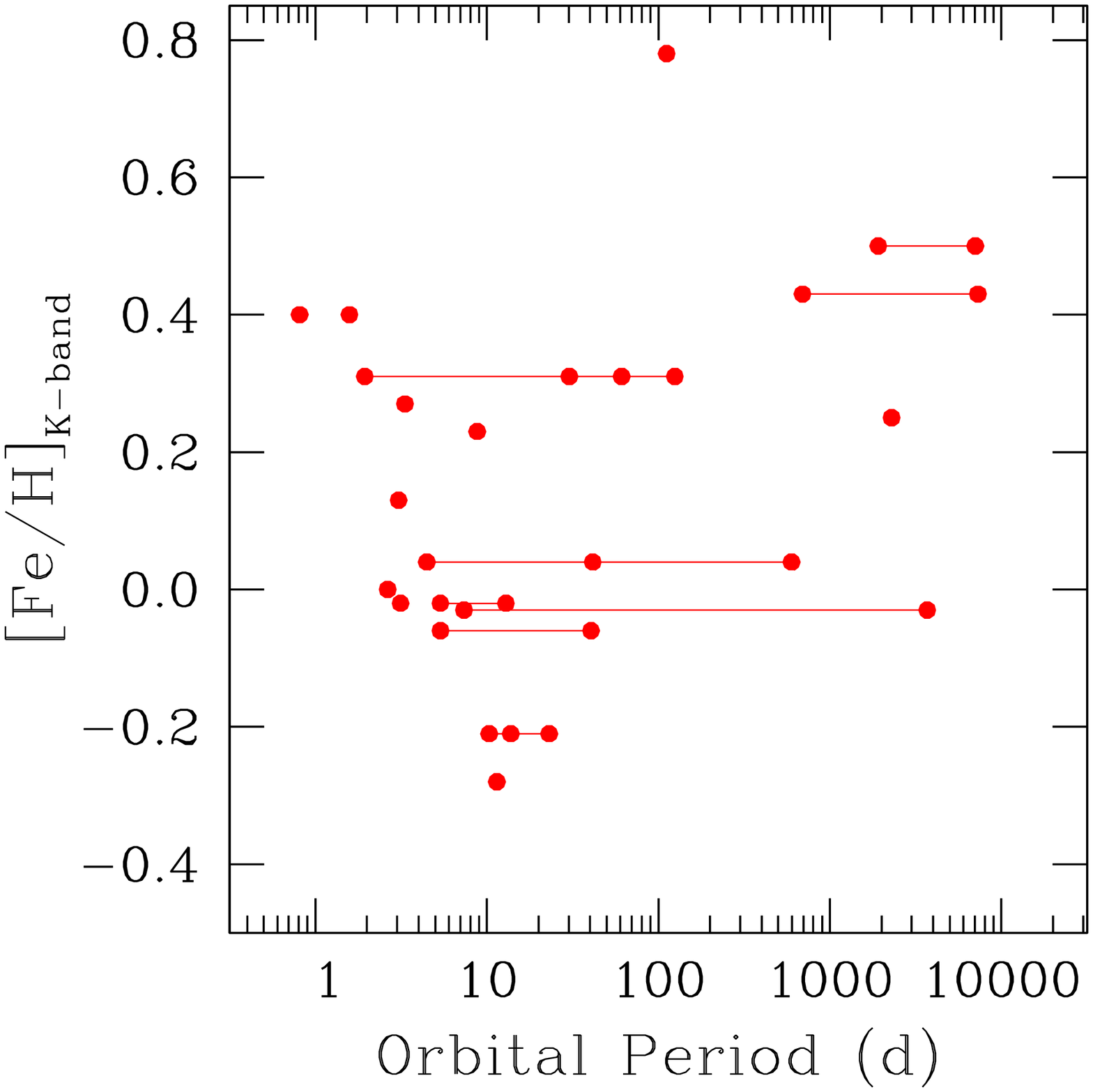}
\includegraphics[width=0.34\columnwidth]{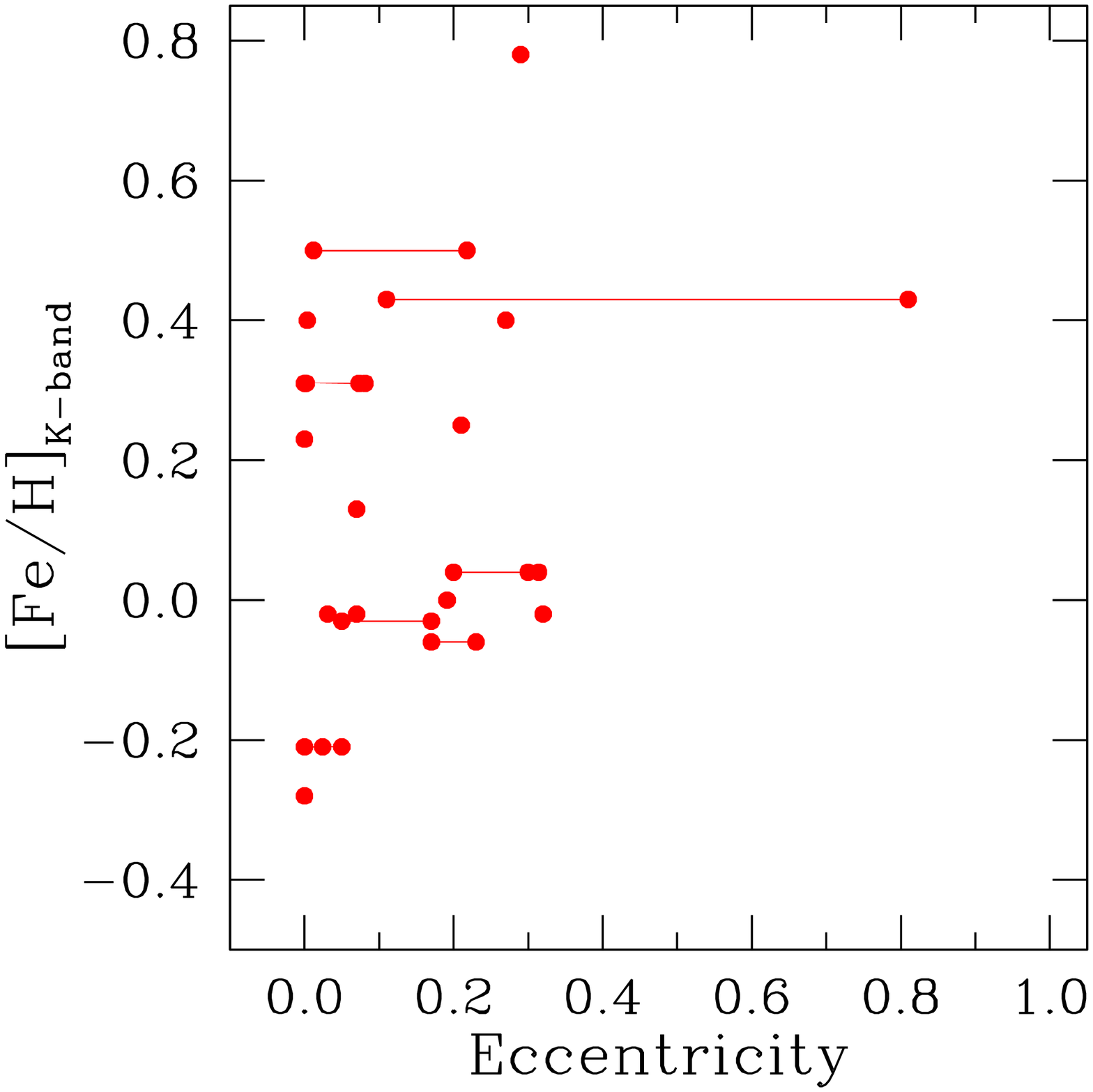}
\caption{K-band metallicity vs period (left panels) and
eccentricity (right panels) for the 11 M-dwarfs with planets observed with GNIRS (upper panels)
and the 18 M-dwarfs planet hosts from the \citet{T15}'s catalogue (lower panels).
Horizontal lines join planets in multiple systems.}
\label{correlations}
\end{center}
\end{figure}
\end{landscape}

Figure \ref{correlations} shows planetary periods (left panels) and eccentricities (right panels) vs 
metallicity for the 11 M-dwarfs observed with GNIRS (upper panels) and the 18 M-dwarfs with planets
in the \citet{T15}'s catalogue (lower panels).   
The orbital period does not show any significant correlation with stellar metallicity.
For FGK stars, \citet{FischerValenti} also found no correlation
between orbital period and stellar metallicity. However, \citet{Adib} found that planets
orbiting metal-poor FGK stars have longer periods than those orbiting metal-rich stars;
they explain this by assuming that planets formed in metal-poor disks form farther out
and/or later and so do not migrate as far in as those from metal-rich disks. While the
apparent lack of such a differentiation may hint at different migration scenarios for
planets around M-dwarfs to those around FGK stars, the low number of objects analysed
here in comparison with those analysed by \citet{Adib} means this must be treated with caution.
The eccentricity does not show any apparent correlation with stellar metallicity, as can be seen
in Figure \ref{correlations}, right panels. This is consistent with both \citet{FischerValenti}'s and
\citet{Adib}'s results for FGK stars, although, again, we should caution about the relatively small
number of objects analysed. 

Finally, it is fair to caution that multiple factors such as stellar temperature may
affect metallicity determinations, particularly for late spectral type stars. As discussed in
the previous sections, on average, M-dwarfs with planets are metal-rich with respect to M-dwarfs
without known planets, providing support to the core accretion model. However, the metallicity
excess is of about 0.10 dex, i.e., not large enough to safely ignore any bias or uncertainty in
the determinations. In the same sense, correlations (or lack thereof) with planetary parameters
should be taken under the caveat of effects that may compromise metallicity determinations
available up to today as well as the relatively small sample of M-dwarfs with planets.

\section{Summary and Conclusions} \label{conc}

In this work, we have determined the effective temperature, radius, luminosity, and metallicity for
a sample of 16 M-dwarf stars (including 11 with planets), using nIR spectra obtained with the GNIRS
instrument on the Gemini North telescope. Metallicities were derived using the calibrations defined by
\citet{RA12},  \citet{T12}, and \citet{Mann13} whilst the remaining stellar parameters were determined
employing the calibrations of \citet{Newton}. In general, for all the parameters we found good agreement,
within error bars, between our values and other determinations from the literature; in this way,
we have shown that GNIRS spectra can be used for the determination of reliable stellar parameters
for M-dwarf stars, and in particular of metallicities.

We adopted  metallicities obtained from the K-band \citet{Mann13}'s calibration and compared the
distributions of metallicities of M-dwarfs with and without known planets in our sample.  
The distributions are significantly different, with stars with planets
being more metallic than those without planets. This result is supported by the analysis of a larger
sample of M-dwarfs with planets (18 stars) and without known planets (213 stars) obtained from the
catalogue of \citet{T15}.

We searched for correlations between the planetary masses, periods and eccentricities and the 
metallicities, using our GNIRS sample of 11 M dwarfs with planet/s and the relatively larger sample
(18 objects) of \citet{T15}. The results coincide, confirming the initial trend derived from GNIRS
spectra.  We found that the sub-sample of M dwarfs with at least one Jupiter-mass planet 
is more metal-rich than the sub-sample with Neptune or super-Earth planets. 
The latter sample is also indistinguishable from the control field.
More metallic stars host larger (giant) planets. However, it must be noted that
for two-thirds of the planets, only the lower bound of $M sin(i)$ is known, not the actual mass.
The planet-metallicity correlation as well as the trend of more metallic stars to host giant planets
support the core accretion model of planetary formation.

In summary, our results suggest that M dwarfs hosting planets follow the
planet-metallicity correlation already observed for FGK stars, as well as
the trend of more metallic stars to host giant planets. In addition, our analyses
show the utility of GNIRS spectra to derive reliable stellar parameters for M dwarfs.
In future, we expect to increase the initial observed sample in order to confirm, with higher
statistical significance and in a homogeneous way, the planet-metallicity correlation of M dwarfs
with planets and the trend of giant planets to preferentially occur around metal-rich stars.

\paragraph{Acknowledgements}
Based on observations obtained at the Gemini Observatory, which is operated by the Association
of Universities for Research in Astronomy, Inc., under a cooperative agreement with the NSF on
behalf of the Gemini partnership: the National Science Foundation (United States), the
National Research Council (Canada), CONICYT (Chile), Ministerio de Ciencia, Tecnolog\'{i}a e
Innovaci\'{o}n Productiva (Argentina), and Minist\'{e}rio da Ci\^{e}ncia, Tecnologia e
Inova\c{c}\~{a}o (Brazil). This research has made use of the SIMBAD database, operated at CDS,
Strasbourg, France. We gratefully acknowledge financial support from CONICET (Consejo Nacional
de Investigaciones Cient\'ificas y T\'ecnicas, Argentina) through grant PIP CONICET
No. 11220120100497. E.J. and R.P. acknowledge the financial support from CONICET in the form
of Post-Doctoral Fellowships. We thank Rachel Mason and Omaira González-Martín for making their
data reduction pipeline publicly available.  We also thank the referee for constructive comments
and suggestions that greatly improved the paper.

\bibliography{biblio}

\end{document}